\documentclass[12pt]{article}
\usepackage[dvips]{graphicx}
\usepackage{amsmath}
\usepackage{graphicx}
\usepackage{amsfonts}
\usepackage{amssymb}
\usepackage{epsfig}
\usepackage{subfigure}

\begin{document}

\title{\bf Effects of Chameleon Scalar Field on Rotation Curves of the Galaxies}
\author{Piyabut Burikham$^{1,2}$\thanks{Email:piyabut@gmail.com}\hspace{1mm} and Sirachak Panpanich$^{1}$\thanks{Email:sirachak.p@student.chula.ac.th,sirachakp@gmail.com} \\
$^1$ {\small {\em  Theoretical High-Energy Physics and Cosmology Group, Department of Physics,}}\\
{\small {\em Faculty of Science, Chulalongkorn University, Bangkok10330, Thailand}}\\
$^2$ {\small {\em  Thailand Center of Excellence in Physics, CHE,
Ministry of Education, Bangkok 10400, Thailand}}\\} \maketitle

\begin{abstract}
We investigate the effects of chameleon scalar field to the
effective density and pressure of a dark matter halo.  The pressure
is generated from the chameleonic fifth force on the matter.  We
demonstrate that the thick-shell non-singular boundary condition
which forbids singular point leads to extremely stringent constraint
on the matter-chameleon coupling when applied to galaxy. We argue
that chameleon profile with central singularity is more likely to
develop in general physical situation. The chameleonic fifth force
from the chameleon profile with central singularity experienced by
the dark matter could significantly modify the rotation curve of
galaxies. The chameleonic fifth force could generate steeper cusp to
the rotation curves in any dark matter profiles starting from the
Navarro-Frenk-White~(NFW)
 to the pseudo-isothermal~(ISO) profile.  Upper limits on the coupling
constant between the chameleon and the dark matter are estimated
from observational
 data of the late-type Low-Surface-Brightness galaxies~(LSB).  It is
 in the order of $\beta < 10^{-3}$.
 \vspace{5mm}


\newpage

\end{abstract}

\section{Introduction}

With series of observations in the recent years, we come to realize
that the expansion of our universe is accelerating
~\cite{novae,cobe,wmap}. However, the exact cause of the
acceleration is not determined for certain. The popular explanations
are based on dark energy models ranging from dark energy in the form
of the cosmological constant, quintessence, phantom, to the
scalar-tensor, modified and $f(R)$ gravity models. The chameleon
dark energy is one of the scalar-tensor theories that use scalar
field to drive the accelerated expansion of the universe. This model
is based on the hypothesis that a scalar field couples with matter
via a conformal transformation. As a consequence, mass of the scalar
field depends on the matter density in a significant way. We call
this kind of scalar a chameleon scalar field after its ability to
adapt itself to the environment.  Its mass becomes large when the
matter is abundant and becomes tiny in the low density region.
Therefore it can effectively hide itself from any detection in such
environment\cite{Khoury:2003rn,Brax:2004px,Brax:2004ym,Waterhouse:2006wv}.
The effects of the chameleon scalar field on the earth are
consequently suppressed and the constraints on the fifth force could
be evaded up to the scale of the solar system.

Since chameleon can adapt its mass very well to the environment,
it can evade most gravitational constraints used to constrain
other scalar gravity theories.  In the models where chameleon
interact with photons, the most stringent constraints on the
chameleon coupling to the photon comes from
Sunyaev-Zel'dovich~(SZ) effect in the Cosmic Microwave Background.
  The SZ measurements of the Coma cluster place strong constraints on the photon-chameleon
  coupling.  The bounds are approximately $g_{\text{eff}}
< 10^{-9}-10^{-8}$ GeV$^{-1}$~\cite{dss}.~\footnote{For the
interaction
$\mathcal{L}=-\frac{1}{4}B_{F}(\phi/M)F_{\mu\nu}F^{\mu\nu},
g_{eff}\equiv (\ln B_{F})_{,\phi}(\phi_{\text{min}}/M)$.}  Weaker
constraints come from the Earth bound experiments such as GammeV
which excluded the region $2\times 10^{-7}<g_{\text{eff}}<4\times
10^{-6}$ GeV$^{-1}$~\cite{GV}.

On the other hand, the strongest constraints on the chameleonic
matter coupling comes from particle colliders and it is merely
$\beta < 2 \times 10^{16}$~\cite{ms1,ms2}~(definition of $\beta$
is given in Section 2).  This constraint is simply the bound on
the new physics mass scale typical for any new physics scenario
investigated at the particle colliders~($M_{\text{new physics}}>
100$ GeV).  In contrast to the chameleon-photon coupling, no
Earth-bound experiments to constrain the chameleon-matter coupling
is operating.  Future tests of gravity in space to measure the
variation of the gravitational constant could place certain bounds
on the matter-chameleon coupling~\cite{Khoury:2003aq,Brax:2004qh}.

A useful insight regarding nature of the chameleon is the fact that
even though the chameleon field increases its mass to hide its
gravitational effect, the chameleon undergoes a spatial variation in
doing so.  When a field varies in space, it generates a density and
pressure. The gradient of pressure then induces a pressure gradient
force subsequently.  This pressure gradient force is actually the
fifth force generated by the chameleon-matter
coupling~\cite{Brax:2004ym}, the very interaction responsible for
changing the chameleonic mass according to the environment.  In this
article, we will use the effect of the chameleonic fifth force on
the matter in the dark matter~(DM) halo to establish an extremely
stringent constraint on the chameleon-matter coupling $\beta$.

The stringent constraint on the matter-chameleon coupling from the
chameleonic fifth force is a direct result of the singular chameleon
solution inevitably developed in a sufficiently large massive object
such as the galactic DM halo. Rapidly raising chameleon field near
the singular point at the center induces appreciably large fifth
force to the matter in the galaxy. The chameleonic fifth force
significantly reduces the circular rotation velocity of matters in
the core region. It will make the cusp of rotation curves steeper.
For the late-type Low-Surface-Brightness~(LSB) galaxies, the
dominant gravitating element is the DM halo.  We use recent
observational data on the rotation curves of the LSB galaxies to
place constraints on the chameleon-matter coupling.  The upper
bounds could be as low as $\beta < 1\times 10^{-3}$ depending on
which type of DM profile is used.

This article is organized as the following.  In Section 2, we review
briefly on the chameleon model and derive the equation of motion
governing its profile in the presence of matter.  In Section 3, we
discuss the inevitability of the chameleon profile with singular
point at the center of the mass distribution.  Enforcing the
non-singular boundary condition leads to extremely stringent
constraint on the matter-chameleon coupling $\beta$~(of order
$10^{-7}$). We argue that this boundary condition and consequently
this constraint is physically unreasonable.  The chameleon profile
with central singularity is then numerically obtained for general
situation and the density, pressure and the fifth force in the
presence of a chameleon scalar field are subsequently calculated.
Effects of chameleonic fifth force to the rotation curves of LSB
galaxies are explored in Section 4 and 5. Constraints on the upper
bound of the matter-chameleon coupling constant are obtained and
discussed in Section 5.  Changes of the power index in the power-law
self potential of chameleon are shown not to affect most results.
Section 6 concludes our work.

\section{Effective Density, Pressure and the Chameleonic Fifth Force}

A chameleon scalar field interacts with matter through a
conformal coupling which could be absorbed into the matter action
as the following~\cite{Khoury:2003rn},
\begin{eqnarray}
S & = & \int d^{4}x\sqrt{-g} \left[
\frac{M^{2}_{Pl}}{2}R-\frac{1}{2}(\partial \phi)^{2}-V(\phi)
\right]-\int d^{4}x \mathcal{L}_{matter}(\tilde{g}_{\mu
\nu},\psi_{m}),
\end{eqnarray}
where $\tilde{g}_{\mu\nu}=A^{2}(\phi)g_{\mu\nu}$, $A(\phi) =
e^{\beta\phi/M_{Pl}}$ is a conformal coupling.

In order to find the effective density and pressure, we need to
determine the profile of the chameleon in dark matter~(DM) halo.
 We set the dynamics of the chameleon in the
thick-shell regime~\cite{Khoury:2003rn,Tamaki:2008mf}.
 Namely, we assume that the value of scalar field which minimizes the effective potential~($\phi_{min}$)
only stays at the exterior of the dark matter halo. The value of
the scalar field in
 the interior of the halo is not $\phi_{min}$ but $\phi(r)$ which is determined
 by the matter density of the halo.
The equation of motion of the chameleon scalar field is
\cite{Brax:2004px}
\begin{equation}
\nabla^2\phi = V,_{\phi}+\alpha_{\phi}\rho_{m}A\left(\phi\right),
\label{eom}
\end{equation}
where $\phi$ is the chameleon scalar field, $\rho_m$ is the matter
density, and $\alpha_{\phi} \equiv \frac{\partial \text{ln}
A(\phi)}{\partial \phi}$.  We can use available density profiles
of dark matter such as the
Navarro-Frenk-White~(NFW)~\cite{Navarro:1995iw} and the
pseudo-isothermal~(ISO) profile~\cite{iso} to simulate the effects
of the chameleon pressure to the DM halo.

For simplicity, we assume that the dark matter halo has spherical
symmetry and the chameleon scalar field is static.  The effect of
spacetime curvature for the DM halo on the chameleon profile is
negligible.  Therefore, Eqn.\eqref{eom} is reduced to
\begin{equation}
\frac{\partial^2 \phi}{\partial r^2} + \frac{2}{r}\frac{\partial
\phi}{\partial r} = V,_{\phi}+\alpha_{\phi}\rho_{m}(r)
A\left(\phi\right).\label{eom1}
\end{equation}

First we will consider the scalar self potential in the form of the inverse-power-law potential $V(\phi) =
\frac{M^{4+n}}{\phi^n}$ \cite{zws,swz,Khoury:2003rn,Brax:2004px}.  Even though this form of self potential is extremely constrained if not already ruled out by the Lunar Laser Ranging~(LLR) experiment and cosmological constraints~Ref.~\cite{gmmptw}, it contains minimal amount of parameters and serves as the simplest chameleon model.  A more viable self potential of the form $V(\phi)=M^{4}(1+\mu(M/\phi)^{n})$ which has not been ruled out by the same analyses, having one extra parameter $\mu$, will be compatible with our result with the substitution $M^{4+n}\to \mu M^{4+n}$~(since only $V,\phi$ appears in the equation of motion).  With the power-law self potential, the
equation of motion becomes
\begin{equation}
\frac{\partial^2 \phi}{\partial r^2} + \frac{2}{r}\frac{\partial
\phi}{\partial r} = -n\frac{M^{4+n}}{\phi^{n+1}} +
\frac{\beta}{M_{Pl}}\rho_{m}(r)
e^{\beta\phi/M_{Pl}}.  \label{eom2}\\
\end{equation}

The right-hand side of the equation can be defined to be the
derivative with respect to the chameleon field of an effective
potential
\begin{eqnarray}
V_{eff} & = & V + \rho_{m}(r) e^{\beta\phi/M_{Pl}}.
\end{eqnarray}
The value of chameleon $\phi_{min}$ which gives minimum effective
potential is then given by
\begin{eqnarray}
\phi_{min}& = & \left( \frac{n M^{4+n} M_{pl}}{\rho_{m}
\beta}\right)^\frac{1}{n+1}.  \label{phimin}
\end{eqnarray}
For our purpose, we will define $\phi_{min}$ for $\rho_{m}=
\rho_{\infty}\text{(average density of the universe})\simeq
10^{-26}$ kg/m$^{3}$.

To consider gravitational effects of the DM halo and the chameleon,
we solve the Einstein equations for a spherically symmetric metric
\begin{equation}
ds^2 =
-A\left(r\right)dt^2+B\left(r\right)dr^2+r^2d\theta^2+r^2\sin^2\theta
d\phi^2.  \label{spherical sym metric}
\end{equation}
The Einstein equations are
\begin{eqnarray}
\frac{B-1}{Br^2} + \frac{B'}{B^2 r} &=& -8\pi G T^{t}_{t}, \\
\frac{B-1}{Br^2} - \frac{A'}{rAB} &=& -8\pi G T^{r}_{r}.
\end{eqnarray}

The energy-momentum tensor in the Einstein equations are the total
energy-momentum tensor $(T^{\mu\nu}_{(total)} =
T^{\mu\nu}_{(matter)} + T^{\mu\nu}_{(\phi)})$ due to the coupling
with matter.  From action of the chameleon scalar field , we
obtain
\begin{equation}
T^{\mu (\phi)}_{\nu} = \partial^{\mu}\phi\partial_{\nu}\phi - \delta^{\mu}_{\nu}\left(\frac{1}{2}g^{\alpha\beta}\partial_{\alpha}\phi\partial_{\beta}\phi+V(\phi)\right)
\end{equation}
and
\begin{eqnarray}
T^{t}_{t (\phi)} &=& -\rho_{(\phi)} = -\frac{\phi^{\prime 2}}{2 B} -  \frac{M^{4+n}}{\phi^n}, \\
T^{r}_{r (\phi)} &=& P^{r}_{(\phi)} = \frac{\phi^{\prime 2}}{2 B} -
\frac{M^{4+n}}{\phi^n}, \\
T^{\theta}_{\theta (\phi)} &=& T^{\phi}_{\phi} = -\frac{\phi^{\prime
2}}{2 B} - \frac{M^{4+n}}{\phi^n}=P^{\theta}=P^{\phi}.
\end{eqnarray}
This is an isotropic distribution with curious behaviour which needs
extra caution since $P^{r}$ and $P^{\theta}, P^{\phi}$ are not
necessarily equivalent~\footnote{In terms of the pressure gradient
force when reduced to the Euler equation of the chameleonic fluid,
there will be extra term $\frac{1}{r}(2P^{r}-P^{\theta}-P^{\phi})$
in addition to $\frac{dP^{r}}{dr}$ in the radial direction.  This
term is required in order to obtain the correct fifth force
expression as in Eqn.~(\ref{fiftheq}).}. Substitute into the
Einstein field equations, the metric then has to satisfy
\begin{eqnarray}
\frac{(B-1)}{Br^{2}} + \frac{B^{\prime}}{B^{2}r} &=& 8\pi G \left(\rho_{m}(r) + \frac{\phi^{\prime 2}}{2 B} +  \frac{M^{4+n}}{\phi^n}\right) \label{einstein tt},\\
\frac{(B-1)}{Br^{2}} - \frac{A^{\prime}}{rAB} &=& 8\pi G \left(-P_{m}-\frac{\phi^{\prime 2}}{2 B} + \frac{M^{4+n}}{\phi^n}\right).\label{einstein rr}
\end{eqnarray}
In the equation of motion of the chameleon scalar field, we will
set the pressure of matter to zero by assuming that the matter in
the universe is in the dust form and ignoring the possible
annihilation pressure of dark matter and
such~\cite{Wechakama:2010yy}.  The effective pressure thus only
comes from the scalar field. The effective density and effective
pressure are then
\begin{eqnarray}
\rho_{eff} &=& \rho_{m}(r) + \frac{\phi^{\prime 2}}{2 B} +  \frac{M^{4+n}}{\phi^n}, \\
P^{r}_{eff} &=& \frac{\phi^{\prime 2}}{2 B} - \frac{M^{4+n}}{\phi^n}
\label{peff}
\end{eqnarray}
respectively.  Observe that the contribution of $\beta$ comes in the
determination of the chameleon profile Eqn.~(\ref{eom2}).  The
effective pressure from the chameleon will thus change with varying
$\beta$ through the changes in the chameleon profile in the dark
matter halo.  However, the dynamics of the matter with respect to
the profile of the chameleon can be obtained directly from the
conservation of the energy-momentum tensor
\begin{eqnarray}
\nabla_{\mu} T^{\mu\nu}_{(matter)}+\nabla_{\mu} T^{\mu\nu}_{(\phi)}
& = & 0,
\end{eqnarray}
which leads to
\begin{eqnarray}
\rho_{m} \partial_{t}\vec{v} & = & -\frac{\beta}{M_{Pl}} \rho_{m}
\vec{\nabla}\phi,
\end{eqnarray}
or
\begin{eqnarray}
 \vec{a}& = & -\frac{\beta}{M_{Pl}}\vec{\nabla}\phi, \label{fiftheq}
\end{eqnarray}
by using the equation of motion, Eqn.~(\ref{eom2}), and assuming a pressureless matter~(see also the Appendix~\ref{appa}).  Namely, the chameleon-matter coupling induces the {\it fifth
force} acting onto the matter~(Ref.~\cite{Brax:2004ym}).  This fifth force could
change the rotation curve of the galaxy substantially provided that
the variation of the chameleon is
sufficiently large as we shall see later in Section 4.

\section{Constraints on the matter-chameleon coupling and singular solutions of the chameleon}

In this section, we will demonstrate starting from the thick-shell
regime, that the chameleon profile within a sufficiently large
massive object can satisfy the non-singular boundary and positivity
condition, $\phi'(0)=0, \phi(\vec{r})\geq 0$ only when $\beta \leq
\beta_{\text{max}}$.

For a region with sufficiently large matter density $\rho_{m}$, an
excellent approximation of the chameleon profile can be obtained
analytically. Since $M_{pl}$ is large~($\simeq 2.4\times 10^{18}$
GeV) and $M$ is small~($\simeq 10^{-3}$ eV), we can approximate
$\alpha_{\phi}\rho(r) A(\phi)\simeq \frac{\beta \rho(r)}{M_{pl}}$
and neglect the potential term in the equation of motion to obtain
\begin{eqnarray}
\phi^{\prime}(r)& \simeq & \frac{\beta}{4 \pi M_{pl}} \left(
\frac{M(r)}{r^{2}}
\right)+\frac{1}{r^{2}}(\phi^{\prime}r^{2}\vert_{r=0}).
\label{anaform0}
\end{eqnarray}
For a boundary condition
\begin{equation}
\phi(r_{max})=\phi_{min}, \phi'(r_{max})\equiv \frac{\gamma
\beta}{4\pi M_{pl}r_{max}^{2}},
\end{equation}
the general solution can be written as
\begin{eqnarray}
\phi'(r) & = & \frac{\beta}{4\pi M_{pl}r^{2}}(M(r)-M_{0}+\gamma),
\label{genfor}
\end{eqnarray}
where $\gamma$ represents~(proportional to) the slope of chameleon
profile at the boundary of the mass distribution and $M_{0}$ is the
total mass of the object(e.g. galaxy) at $r_{max}$.  This expression
can be integrated directly for any mass profile $M(r)$ to obtain the
corresponding chameleon solution.

There are three classes of positive solutions categorized by the
value of
$\gamma$, \\
1. $\gamma < M_{0}$ ; singular at $r=0$ \\
2. $\gamma = M_{0}$ ; $\phi^{\prime}r^{2}\vert_{r=0}=0$~(nonsingular) \\
3. $\gamma > M_{0}$ ; truncated at finite $r$~(unphysical). \\

\begin{figure}[htp]
\centering
\includegraphics[width=0.45\textwidth]{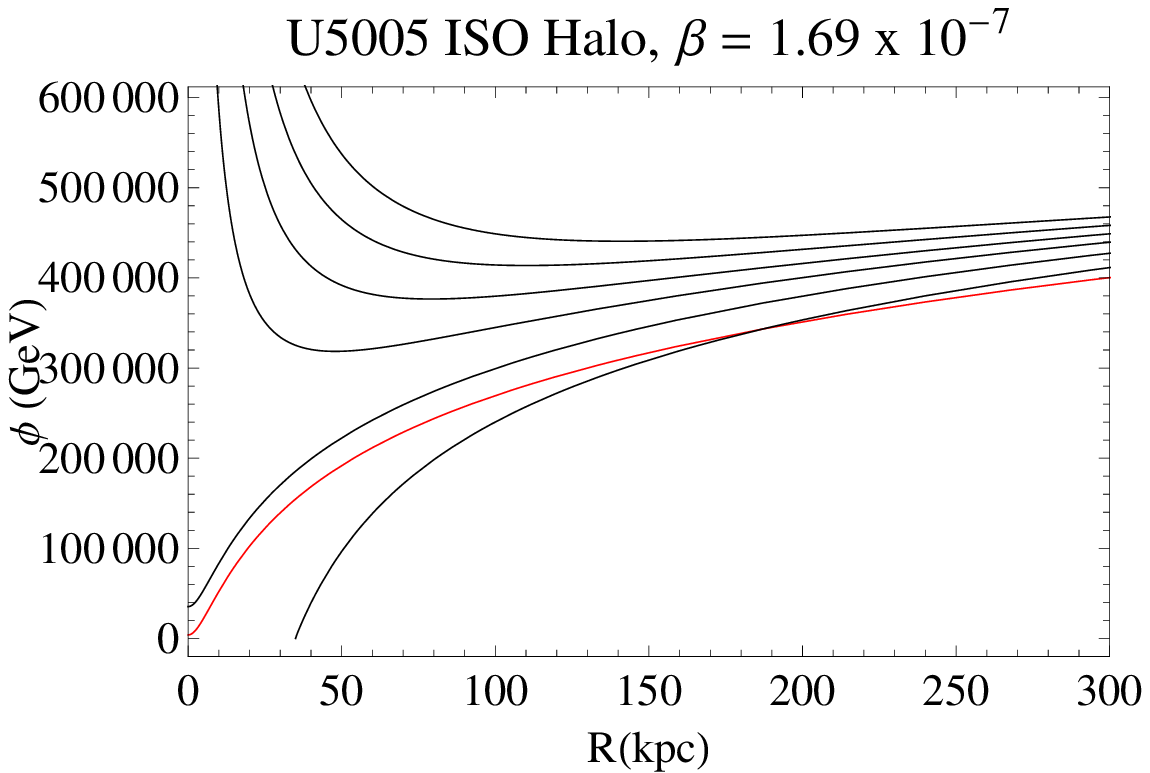} \hfill
\includegraphics[width=0.45\textwidth]{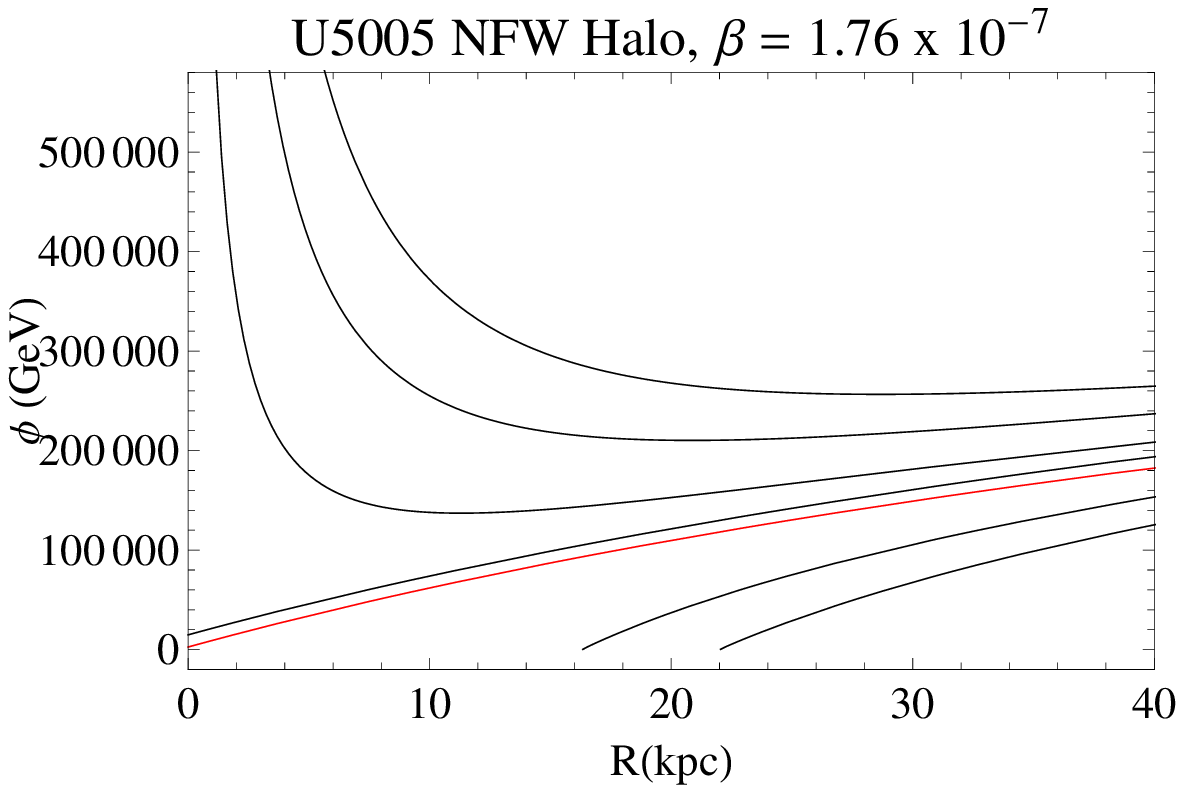}\\
\caption[solutions]{The chameleon solutions for various $\gamma$.
The red curve is the approximated analytic solution neglecting the potential $V(\phi)$.  The ISO solution~(left) satisfies
non-singular boundary condition $\phi'(0)=0$ but not the NFW~(right).  The other curves are
obtained numerically. }\label{figsol1}
\end{figure}

Varying the slope $\gamma > 0$, various chameleon solutions~(using the full equation of motion, Eqn.~(\ref{eom2}), with the potential included) for
sufficiently small $\beta$ can be obtained numerically as are shown
in Fig. \ref{figsol1}.  All three kinds of solutions are presented.

The third kind of solution is unphysical since it is truncated at
finite radial distance.  Among the remaining physical solutions, the
only nonsingular one satisfies the boundary condition $\phi'r^{2}\vert_{r=0}=0$~(class 2).  The nonsingular solutions of the chameleon are investigated by most literature with the boundary condition $\phi'(0)=0$ and finite $\phi(0)$.  In the following subsection, we will demonstrate that for certain matter profiles, this boundary condition cannot be satisfied.  And for those matter profiles, the most natural chameleon solution would be the singular one.

\subsection{Analytic solutions for small potential and remarks on thick-shell boundary conditions} \label{appb}

In this subsection, we consider analytic solutions of the chameleon coupling to various DM models: the NFW, ISO, and the parametrized model~(PM).  The potential term $V(\phi)$ will be neglected.  Subsequently, both the approximated analytic and numerical~(with the potential included) solutions will be compared and justified the validity of the approximation.  We will investigate whether which DM profile allows non-singular chameleonic solution with flat boundary condition $\phi^{\prime}(0)=0$.

For nonsingular case in class 2, we have the gradient of the scalar given in terms of the matter density as
\begin{eqnarray}
\phi^\prime r^2 &=& \frac{\beta}{M_{Pl}} \int \rho(r) r^2 dr, \label{phip}
\end{eqnarray}
where $\rho(r)$ is the matter distribution.

\begin{figure}[htp]
\centering
\includegraphics[width=5in]{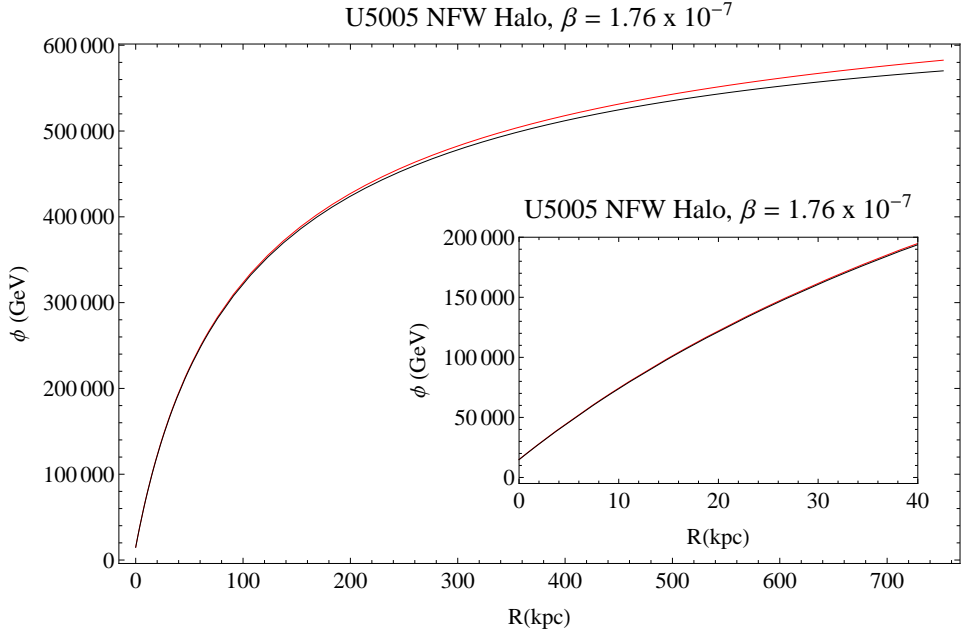}
\caption[profile]{The chameleon profile in the NFW DM halo of
U5005 galaxy for $\beta = 1.76\times10^{-7}$. The analytic and numerical chameleon solutions are represented in red line and black line respectively. }\label{figpp1}
\end{figure}

For the NFW profile : $\displaystyle{\rho_{NFW}(r) = \frac{\rho_0}{\frac{r}{a}(1+\frac{r}{a})^2}}$.
Substitute into Eqn.~(\ref{phip}) gives
\begin{eqnarray}
\phi^\prime r^2 &=& \frac{a^3 \beta \rho_0}{M_{Pl}} \left(\frac{a}{a + r} + \text{ln}(a + r)\right) + C_1.
\end{eqnarray}
Taking limit $r \rightarrow 0$ and assuming $\phi^{\prime}\to \infty$ slower than $1/r^{2}$ in this limit, the constant $C_{1}$ can be determined to be
\begin{eqnarray}
 C_1 &=& - \frac{a^3 \beta \rho_0}{M_{Pl}} \left(1 + \text{ln}(a)\right).
\end{eqnarray}
Therefore
\begin{eqnarray}
\phi^\prime r^2 &=& \frac{a^3 \beta \rho_0}{M_{Pl}} \left(\frac{a}{a + r} + \text{ln}(a + r)\right) - \frac{a^3 \beta \rho_0}{M_{Pl}} \left(1 + \text{ln}(a)\right),\nonumber\\
\phi^\prime(r) &=& \frac{a^3 \beta \rho_0}{M_{Pl} r^2}\left(\text{ln}(1 + \frac{r}{a}) - \frac{r}{a+r}\right).
\end{eqnarray}
This is the solution of $\phi^\prime$ for arbitrary initial condition, $\phi^\prime(r = 0)$.  The solution of the chameleon scalar field in the NFW dark matter halo can then be obtained,
\begin{eqnarray}
\phi(r) &=& \frac{a^3 \beta \rho_0}{M_{Pl}} \left(\frac{1}{a} - \frac{\text{ln}(1 + \frac{r}{a})}{r}\right) + \phi(0).
\end{eqnarray}
Thus, we can see that at the origin of the NFW dark matter halo $(r = 0)$ the chameleon profile cannot be flat with
\begin{eqnarray}
\phi^\prime(0) = \frac{a \beta \rho_0}{2 M_{Pl}}.
\end{eqnarray}
This proves conclusively that for the chameleon coupling to the NFW DM in the thick-shell regime, the chameleon field cannot satisfy the boundary condition $\phi^\prime(0)=0$.  Figure~\ref{figpp1} shows both the approximated analytic and numerical~(with potential included) solutions, the difference is minimal.

\begin{figure}[htp]
\centering
\includegraphics[width=5in]{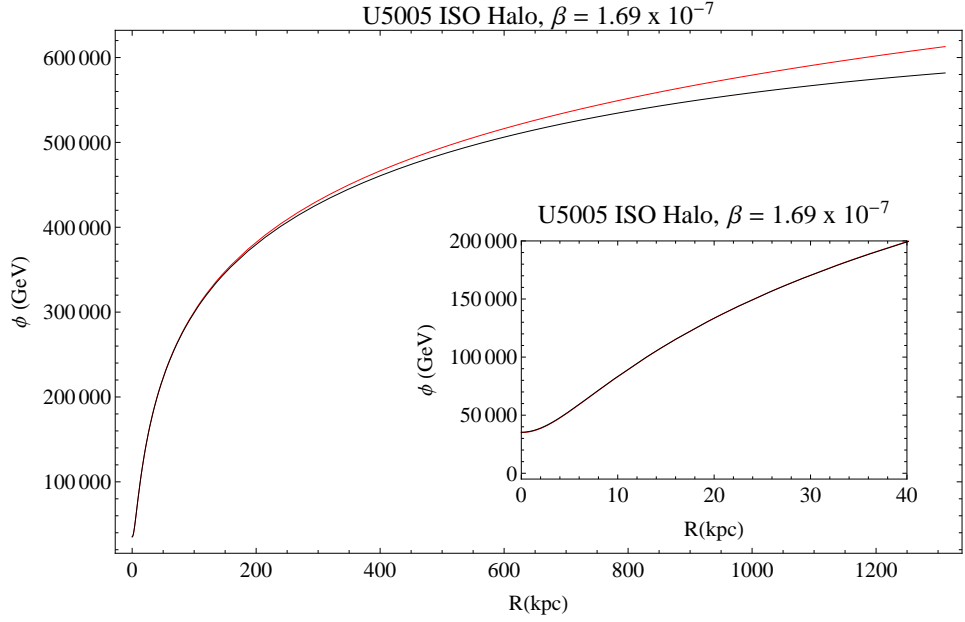}
\caption[profile]{The chameleon profile in the ISO DM halo of
U5005 galaxy for $\beta = 1.69\times10^{-7}$. The analytic and numerical chameleon solutions are represented in red line and black line respectively. }\label{figpp2}
\end{figure}

For the ISO profile: $\displaystyle{\rho_{ISO}(r) = \frac{\rho_0}{1+(\frac{r}{R_s})^2}}$,
\begin{eqnarray}
\phi^\prime r^2 &=& \frac{\beta}{M_{Pl}} \int \rho_{ISO}(r) r^2 dr, \nonumber
\end{eqnarray}
leading to
\begin{eqnarray}
\phi^\prime r^2 &=& \frac{\beta}{M_{Pl}}R_s^3 \rho_0 \left(\frac{r}{R_s}-\arctan(\frac{r}{R_s})\right)+C_1.
\end{eqnarray}
Taking limit $r\rightarrow 0$ and assuming $\phi^{\prime}\to \infty$ slower than $1/r^{2}$ in this limit, we obtain $C_{1}=0$.
Therefore
\begin{eqnarray}
\phi^\prime (r) &=& \frac{\beta R_s^3 \rho_0}{M_{Pl} r^2}\left(\frac{r}{R_s}-\arctan(\frac{r}{R_s})\right).
\end{eqnarray}
Integrate to obtain solution of the chameleon scalar field in ISO dark matter halo,
\begin{eqnarray}
\phi(r) &=& \frac{\beta R_s^3 \rho_0}{M_{Pl}} \left(\frac{\arctan(r/R_s)}{r}+\frac{\text{ln}(1+r^2/R_s^2)}{2R_s}-\frac{1}{R_s}\right)+\phi(0).
\end{eqnarray}
Thus, the chameleon profile at the origin of the ISO dark matter halo can be flat with $\phi^\prime(0) = 0$.  Figure~\ref{figpp2} shows both the approximated analytic and numerical~(potential included) solutions.  The approximation becomes worse as the radial distance grows.

For the parametrized model (PM) : $\displaystyle{\rho_{PM}(r) = \frac{\rho_0}{(\frac{r}{r_s})^{\alpha}(1+\frac{r}{r_s})^{3-\alpha}}}$,
\begin{eqnarray}
\phi^\prime r^2 &=& \frac{\beta}{M_{Pl}} \int \rho_{PM}(r) r^2 dr, \nonumber\\
&=& \frac{\beta}{M_{Pl}} \frac{r^{3-\alpha} r_s^{\alpha}\rho_0 ~{}_2F_1(3-\alpha,3-\alpha,4-\alpha,-\frac{r}{r_s})}{3-\alpha}+C_1,
\end{eqnarray}
where ${}_2F_1$ is the hypergeometric function. \\

Taking limit $r\rightarrow 0$ and assuming $\phi^{\prime}\to \infty$ slower than $1/r^{2}$ in this limit, we obtain $C_{1}=0$.  Then
\begin{eqnarray}
\phi^\prime(r) &=& \frac{\beta}{M_{Pl}} \frac{r^{1-\alpha} r_s^{\alpha}\rho_0 ~{}_2F_1(3-\alpha,3-\alpha,4-\alpha,-\frac{r}{r_s})}{3-\alpha},
\end{eqnarray}
or
\begin{eqnarray}
\phi^\prime(r) &=& \frac{\beta}{M_{Pl}} \frac{r^{1-\alpha} r_s^{\alpha}\rho_0 (1+\frac{r}{r_s})^{\alpha-3}{}_2F_1(3-\alpha,1,4-\alpha,\frac{r/r_s}{1+r/r_s})}{3-\alpha}.
\end{eqnarray}
The value of $\phi^\prime(0)$ for each $\alpha$ can then be obtained,\\ \\
Case I : $\alpha < 1$
\begin{eqnarray}
\phi^\prime(0) = 0.
\end{eqnarray}
Case II : $\alpha = 1$~(NFW)
\begin{eqnarray}
\phi^\prime(0) = \frac{r_s \beta \rho_0}{2 M_{Pl}}.
\end{eqnarray}
Case III : $1 < \alpha < 2$
\begin{eqnarray}
\phi^\prime(0) = \infty.
\end{eqnarray}

\begin{figure}[htp]
\centering
\includegraphics[width=5in]{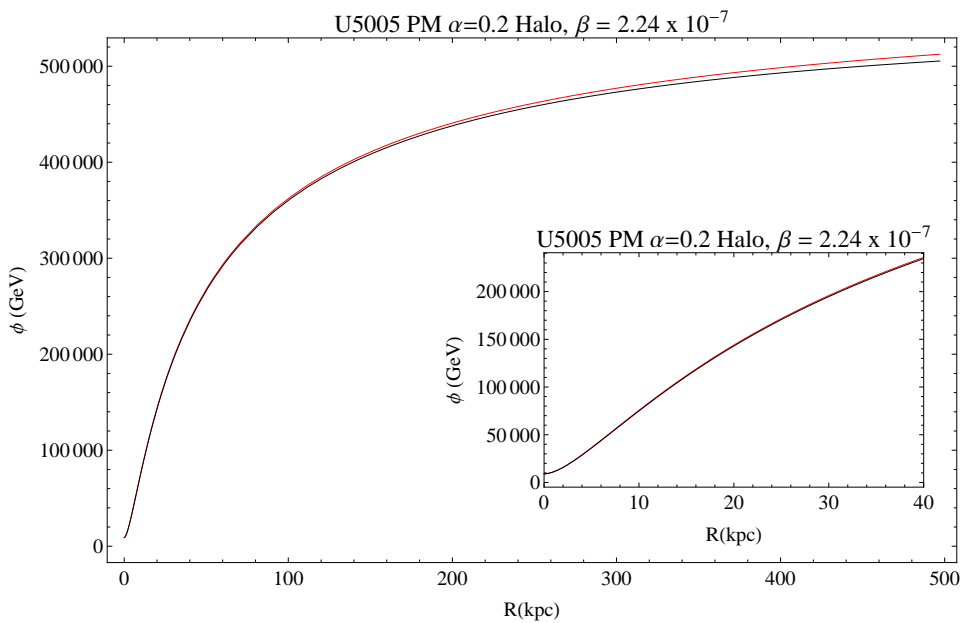}
\caption[profile]{The chameleon profile in the PM DM halo of
U5005 galaxy for $\alpha = 0.2$ and $\beta = 2.24\times10^{-7}$. The analytic and numerical chameleon solutions are represented in red line and black line respectively. }\label{figpp4}
\end{figure}

\begin{figure}[htp]
\centering
\includegraphics[width=5in]{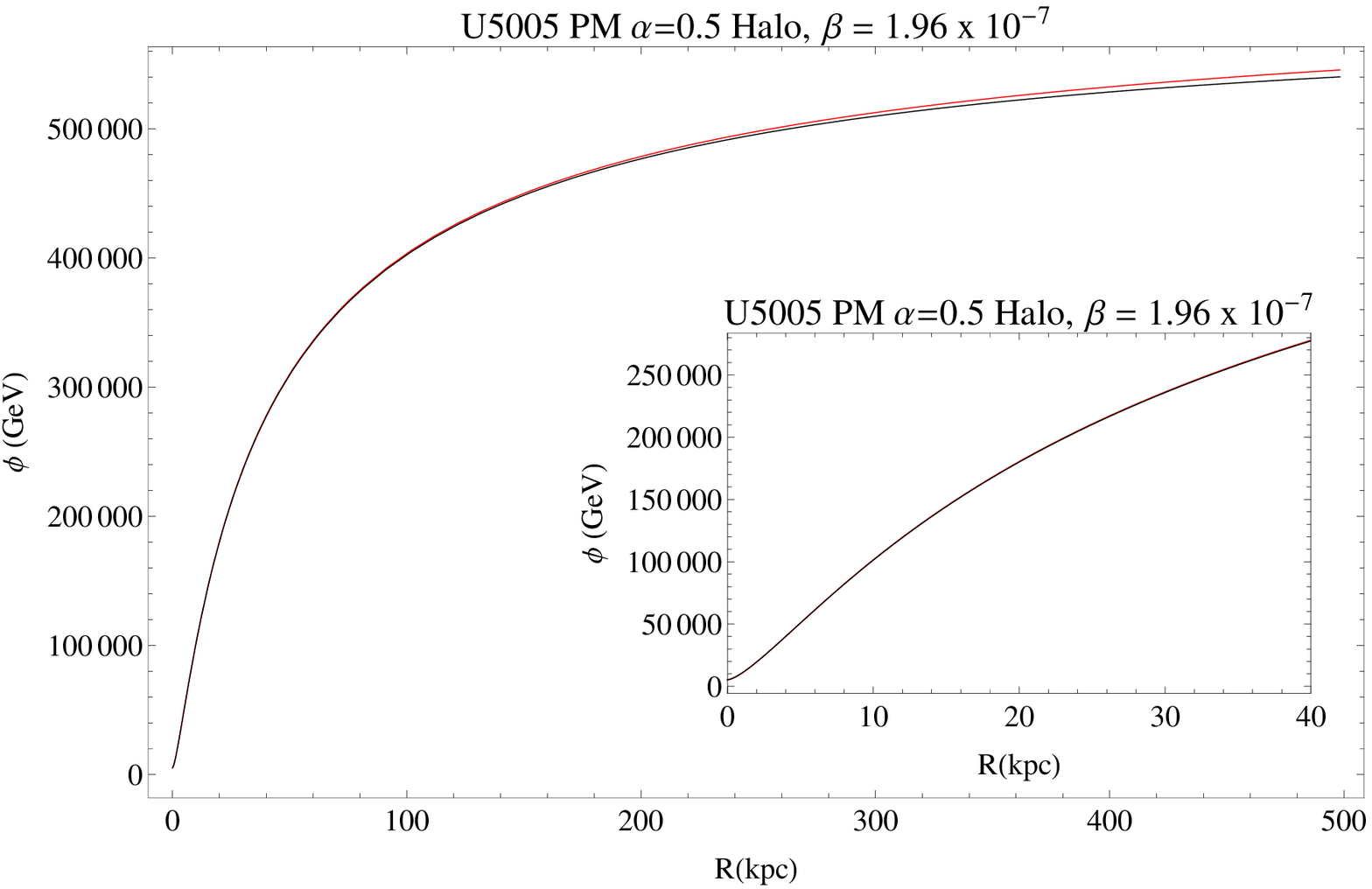}
\caption[profile]{The chameleon profile in the PM DM halo of
U5005 galaxy for $\alpha = 0.5$ and $\beta = 1.96\times10^{-7}$. The analytic and numerical chameleon solutions are represented in red line and black line respectively. }\label{figpp5}
\end{figure}

\begin{figure}[htp]
\centering
\includegraphics[width=5in]{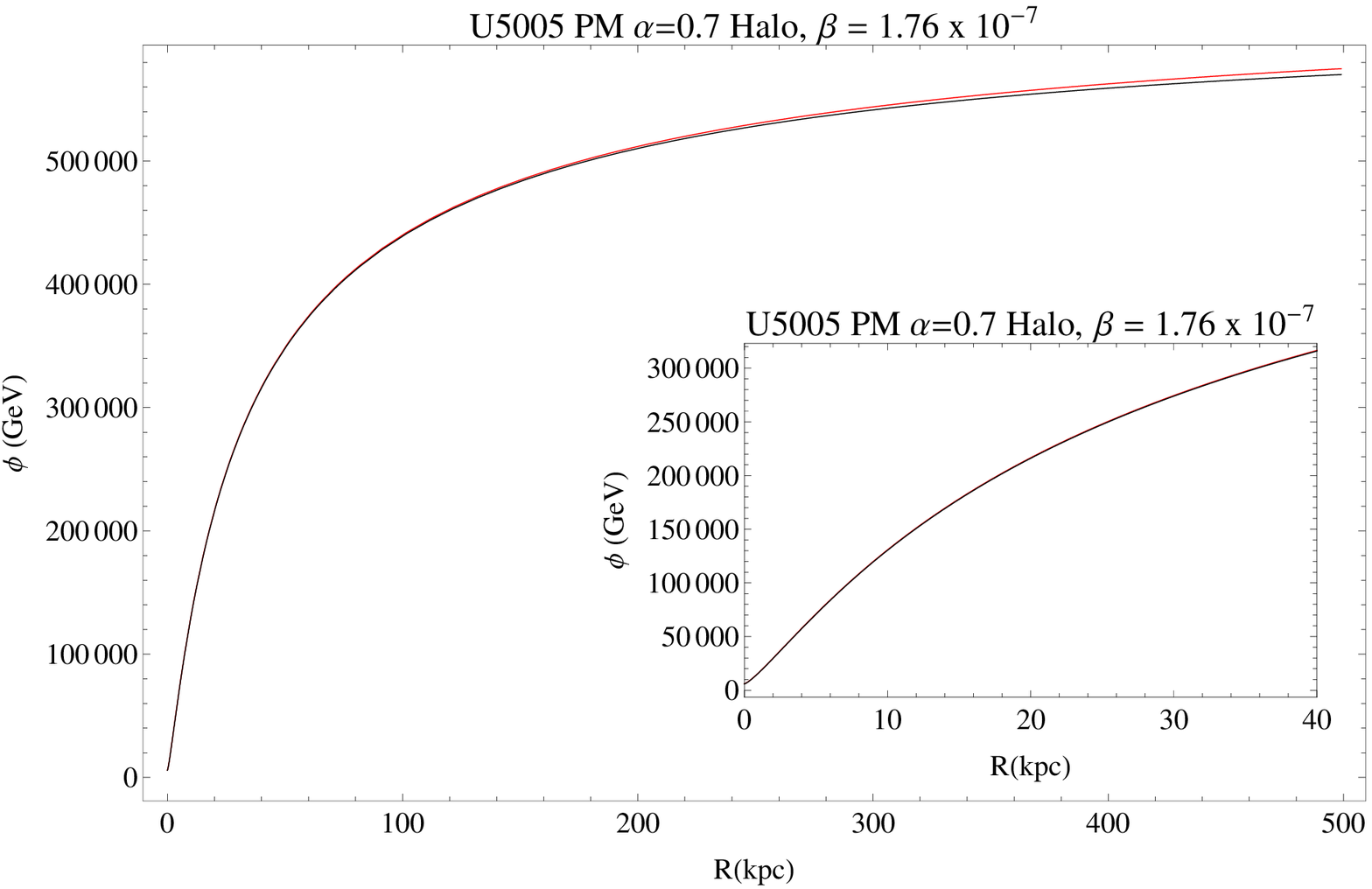}
\caption[profile]{The chameleon profile in the PM DM halo of
U5005 galaxy for $\alpha = 0.7$ and $\beta = 1.76\times10^{-7}$. The analytic and numerical chameleon solutions are represented in red line and black line respectively. }\label{figpp6}
\end{figure}

\begin{figure}[htp]
\centering
\includegraphics[width=5in]{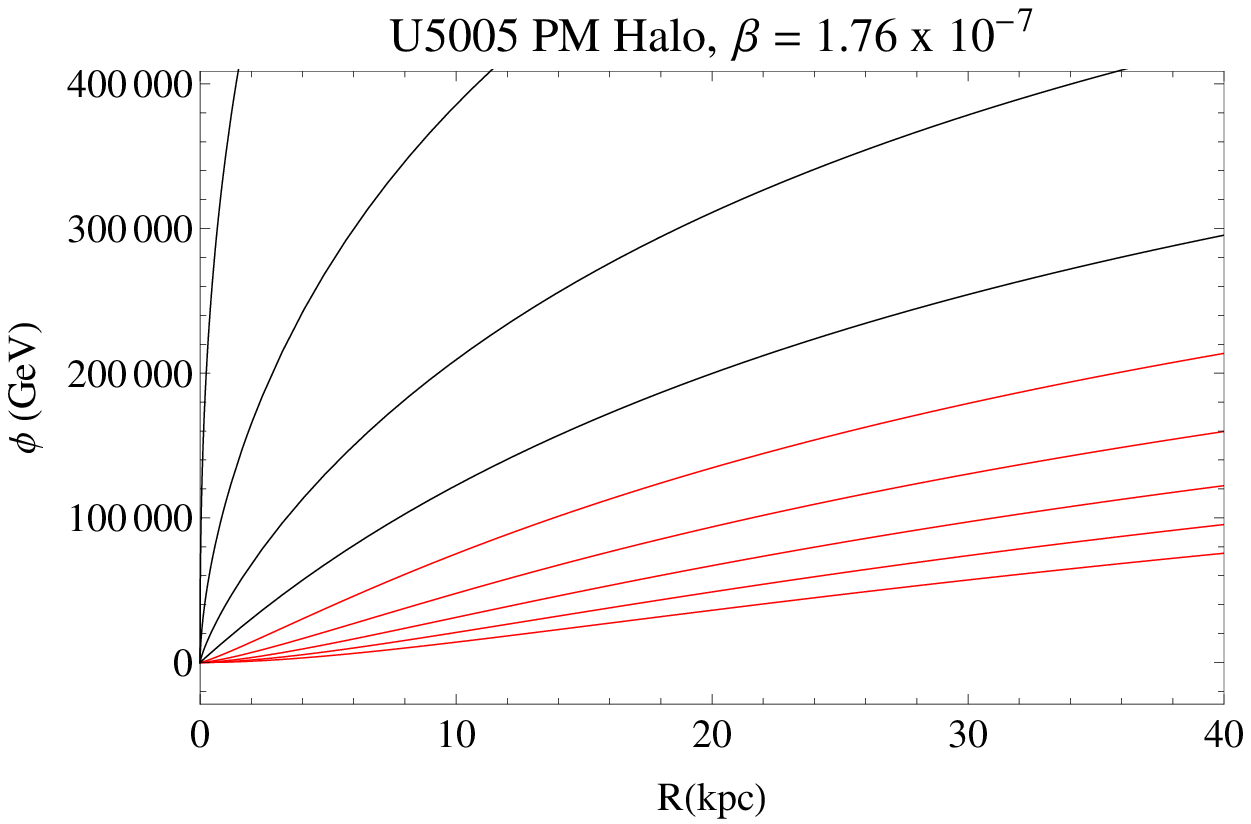}
\caption[profile]{The analytic chameleon profiles in the PM DM halo of
U5005 galaxy for $\beta = 1.76\times10^{-7}$. The red lines represent the chameleon profile for $0<\alpha<1$ and the black lines represent the chameleon profile for $1\leq \alpha < 2$.}\label{figpp3}
\end{figure}

The corresponding solution of the chameleon scalar field in the PM dark matter halo is
\begin{eqnarray}
\phi(r) &=& \frac{\beta \rho_0}{M_{Pl}} \frac{r^{\alpha}}{3-\alpha}\left(\frac{r^{2-\alpha}}{2-\alpha} {}_2F_1(2-\alpha,3-\alpha,4-\alpha,-\frac{r}{r_s})\right)+\phi(0),
\end{eqnarray}
or
\begin{eqnarray}
\phi(r) &=& \frac{\beta \rho_0}{M_{Pl}} \frac{r^{\alpha}}{3-\alpha}\left(\frac{r^{2-\alpha}}{2-\alpha} (1+\frac{r}{r_s})^{\alpha-2}{}_2F_1(2-\alpha,1,4-\alpha,\frac{r/r_s}{1+r/r_s})\right)+\phi(0).
\end{eqnarray}
where $\alpha$ must be less than $2$.  Figure~\ref{figpp4},\ref{figpp5},\ref{figpp6} show both approximated analytic and numerical~(potential included) solutions for the PM DM profiles, the differences are hardly visible.  Figure~\ref{figpp3} summarizes the numerical solutions for the PM model.  The behaviour of the boundary value $\phi^{\prime}(0)$ confirms the analytic results.

As a summary, we have shown that depending on the DM profile, the chameleon solution canNOT arbitrarily satisfy the boundary condition $\phi^{\prime}(0)=0$.  The NFW and the parametrized profiles with $\alpha > 1$~($\alpha$ must be less than 2 for positivity of the chameleon distribution function) cannot satisfy this flat boundary condition as shown above.

Now we want to derive a bound on the coupling $\beta$ for the nonsingular solutions which require $\phi$ at the galactic edge to match $\phi_{min}$ of the universe.  The value of $\phi_{min}$ is determined from the observed value of the dark energy through the self-interacting potential of the chameleon.  Using the analytic formula, we can calculate the maximal
matter-chameleon coupling of this solution by integrating
Eqn.~(\ref{anaform0}),
\begin{eqnarray}
\phi_{min} & \geq & \frac{\beta}{4\pi
M_{pl}}\int^{r_{max}}_{0}dr \frac{1}{r^{2}}M(r). 
\end{eqnarray}
With $\phi_{min}$ given by Eqn.~(\ref{phimin}), the maximum
$\beta_{max}$ is found to be
\begin{eqnarray}
\beta_{max} & = & \left( \frac{n M^{4+n}
M_{pl}}{\rho_{\infty}}\right)^{\frac{1}{n+2}} \left[ \frac{4\pi
M_{pl}}{\int^{r_{max}}_{0} dr \frac{M(r)}{r^{2}}}
\right]^{\frac{n+1}{n+2}}. \label{bmax}
\end{eqnarray}
This crucial limit on the matter-chameleon coupling is originated
from the non-singular boundary condition $\phi'(0)=0$.  As the
object gets more and more massive and substantial, the maximum value
of $\beta$ decreases accordingly.  For galaxy U5005 with ISO and NFW
DM profile, the values of $\beta_{max} \simeq 1.69, 1.76\times
10^{-7}$ respectively.

For the self potential of the form $V(\phi)=M^{4}(1+\mu (M/\phi)^{n})$, we can make substitution $M^{4+n}\to \mu M^{4+n}$ to obtain the limit containing two parameters $\beta, \mu$ as
\begin{eqnarray}
\beta & \lesssim & \left( \frac{n \mu M^{4+n}
M_{pl}}{\rho_{\infty}}\right)^{\frac{1}{n+2}} \left[ \frac{4\pi
M_{pl}}{\int^{r_{max}}_{0} dr \frac{M(r)}{r^{2}}}
\right]^{\frac{n+1}{n+2}}.
\end{eqnarray}
For $n=1$, the value of $\mu$ is constrained by the LLR experiment and cosmological conditions to be smaller than $10^{5}$~\cite{gmmptw}, resulting in the upper limit $\beta \lesssim 10^{5/3}\beta_{max}$ for $\beta_{max}$ given in Eqn.~(\ref{bmax}).  This is roughly 46 times larger than the original inverse-power-law self potential case.  For NFW and ISO DM profile of galaxy U5005, the upper limit corresponds to about $\beta < 10^{-5}$.

Among the three classes of solutions mentioned earlier, only the one satisfying
non-singular boundary condition was previously considered physically
relevant. However, as demonstrated above, the non-singular boundary
condition canNOT be satisfied for arbitrary substantially massive
object such as the galactic DM halo.  At the time of structure
formation when the scalar field obtained their vev as the minimum of
the effective potential, there is no way the chameleon could know
which value the matter-chameleon coupling $\beta$ should be. This
value should have been fixed by some theory at the high scale.  For
generic situation during structure formation, the chameleon should
thus develop a profile with singular point at the center $r=0$.
There is no physical reason to prevent this class of solutions for the galaxy.

\subsection{Chameleon solutions with singular point}

\begin{figure}[htp]
\centering
\includegraphics[width=0.45\textwidth]{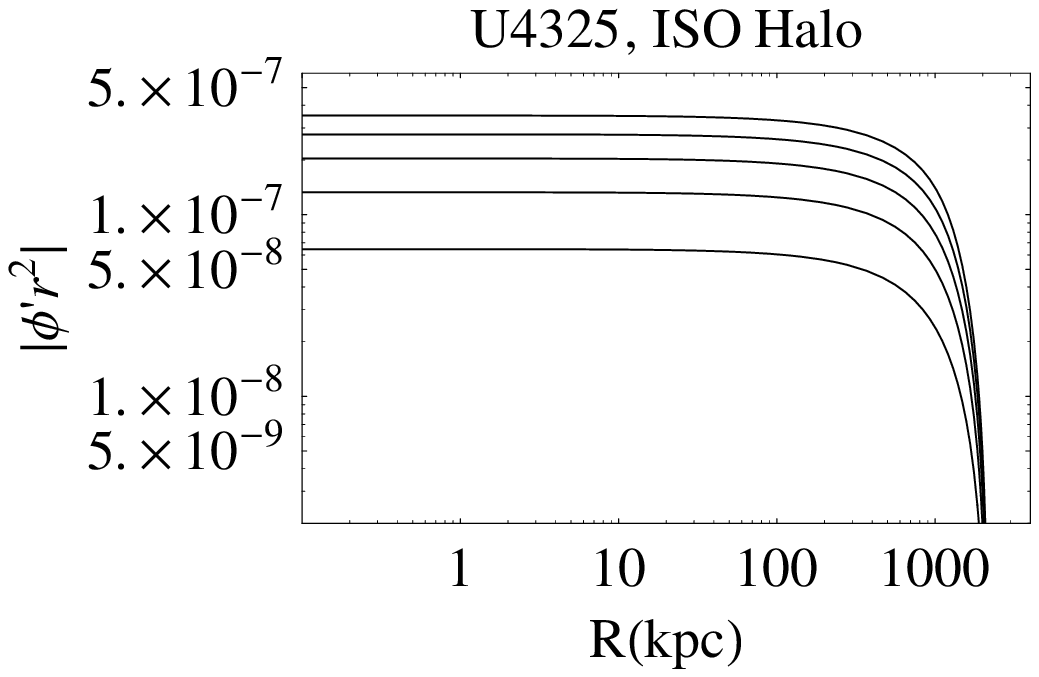} \hfill
\includegraphics[width=0.45\textwidth]{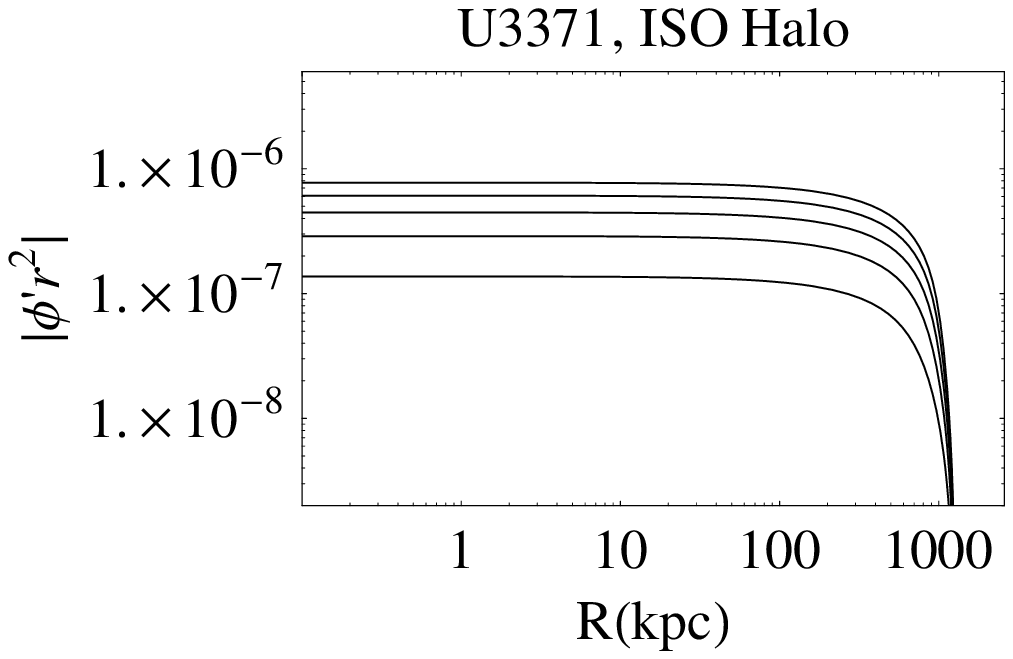}\\
\includegraphics[width=0.45\textwidth]{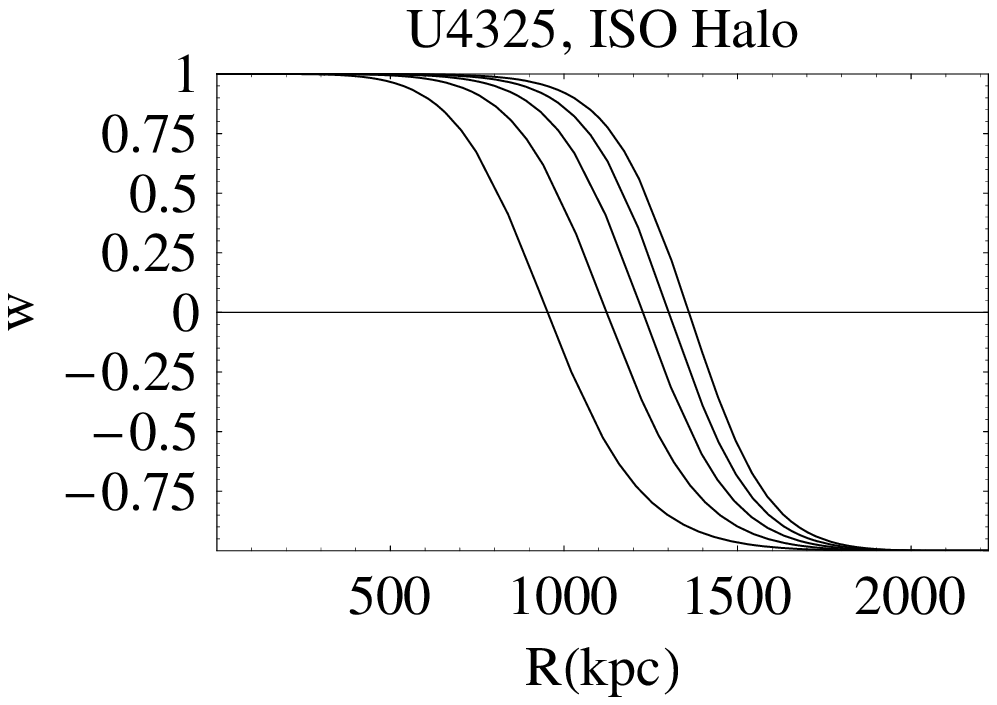} \hfill
\includegraphics[width=0.45\textwidth]{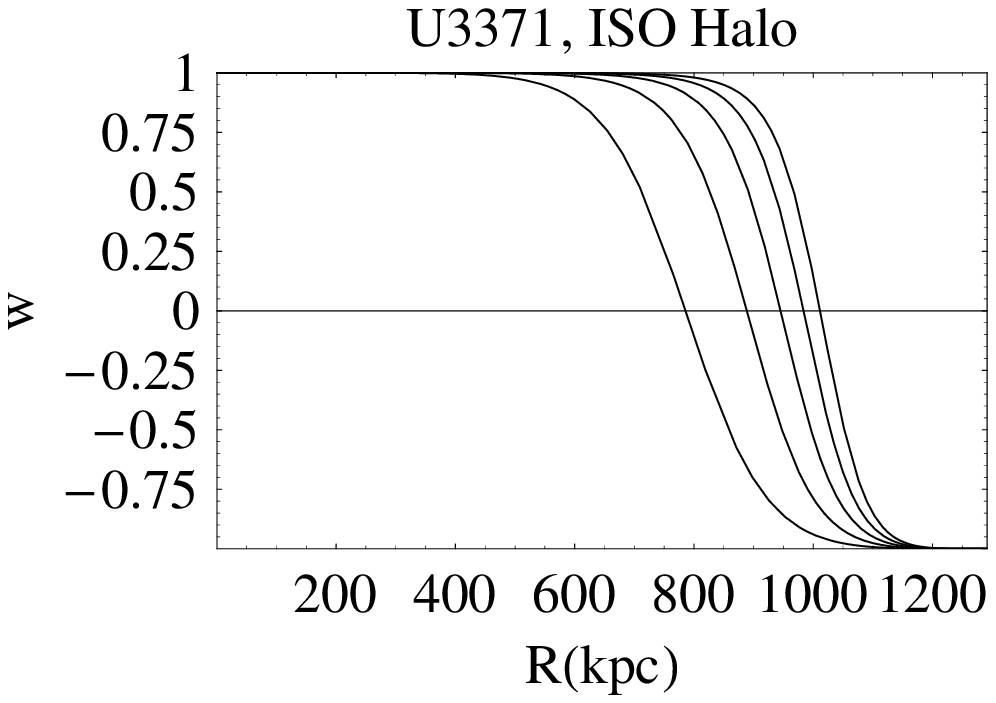}\\
\caption[profile]{The chameleon profile and the ratio
$w=P_{\phi}/\rho_{\phi}$ of the chameleon in the ISO DM halo of
U4325, U3371 galaxy for varying $\frac{\beta}{10^{-7,-6}} =
2,4,6,8,10$~(from bottom to top graph). The chameleon field and
$w$ increase with the coupling $\beta$. }\label{figp1}
\end{figure}

In general situation, the chameleon could develop singular profile
at $r=0$ when $\gamma < M_{0}$.  For sufficiently large $\beta$, the
chameleon solutions with $0< \gamma < M_{0}$ yield mostly the same
results as the case $\gamma = 0$.  Without loss of generality, we
will therefore consider the profile with $\gamma = 0$ and
numerically solve for the profile of chameleon within the dark
matter halo using the boundary condition
$\phi(r_{max})=\phi_{min},\phi^{\prime}(r_{max})=0$ for $r_{max}$
defined to be the radial distance where $\rho_{m}(r_{max})\simeq
\rho_{\infty}$, the average density of the universe.  We will set
$n=1$ for the chameleonic self-interacting power-law potential
$V(\phi)=M^5/\phi$ and briefly discuss the negligible changes for
other $n$ in Section 5. The chameleon profiles for the NFW and ISO
dark matter halo for some LSB galaxies are shown in
Fig.~\ref{figp1}.

\begin{figure}[htp]
\centering
\includegraphics[width=0.45\textwidth]{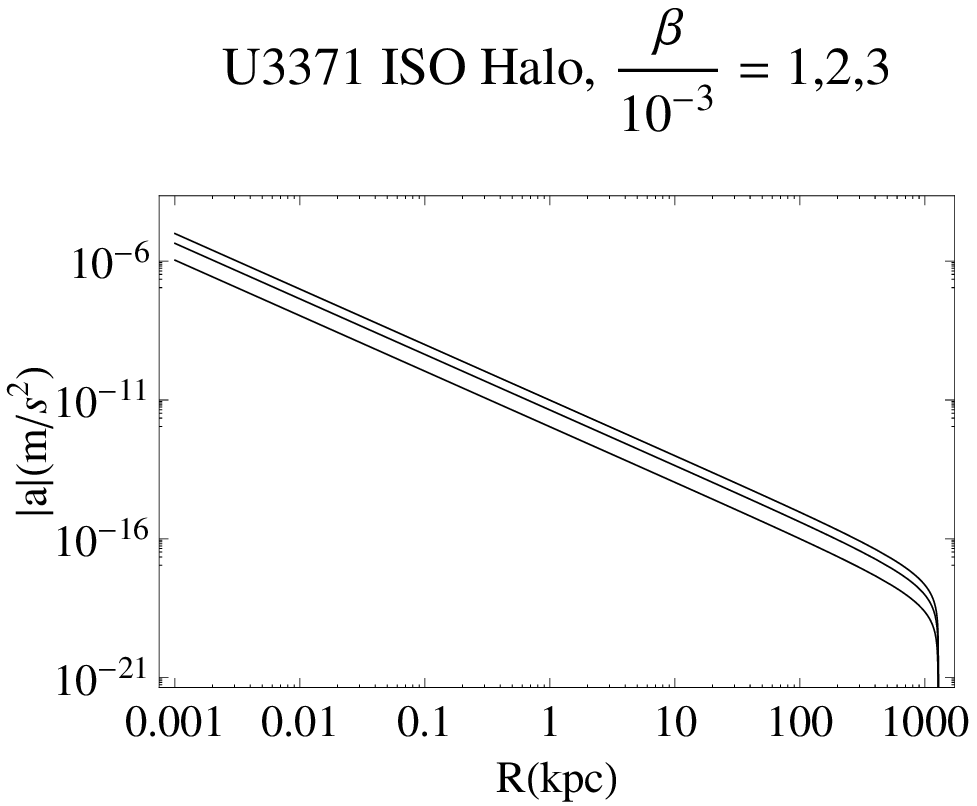} \hfill
\includegraphics[width=0.45\textwidth]{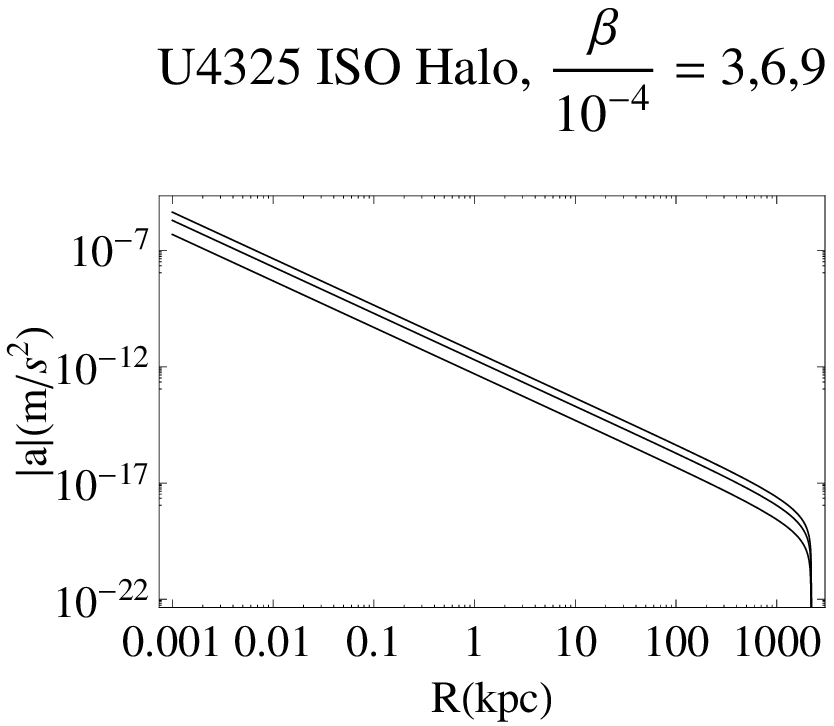}\\
\caption[profile]{The acceleration from the chameleonic fifth force
in the ISO DM halo of U4325, U3371 galaxy. The acceleration
increases with the coupling $\beta$. }\label{figp2}
\end{figure}

The density of chameleon turns out to be only about $10^{-5}$ of the
DM density and thus no distinguishable effects of the chameleon
could be seen in the gravitational lensing.  The gravitational
contribution of the chameleon pressure is also negligible.
Regardless of its smallness as a gravitational entity, the chameleon
pressure gradient induces appreciable force on the DM halo in the
form of the fifth force.

The chameleon changes from dark energy to matter as the density of
the DM increases towards the galactic center.  This is shown by the
ratio $w=P_{\phi}/\rho_{\phi}$ as in Fig.~\ref{figp1}.  As it
changes into some form of exotic chameleonic matter with $w=1$, the
acceleration of the test object due to the chameleonic fifth force
increases as shown in Fig.~\ref{figp2}.

\section{Rotation Curves}

The fifth force induced by the chameleon on the dark matter has
variation along the radial direction according to
Eqn.~(\ref{fiftheq}).  The circular velocity is then reduced when
the gradient of the chameleon is negative. The circular velocity
which includes the acceleration due to the fifth force from the
chameleon can be written as the following
\begin{equation}
v_{c}(r) = \sqrt{\frac{G M(r)}{r} + \frac{\beta
r}{M_{Pl}}\frac{d\phi}{dr}}. \label{rotc}
\end{equation}
The accumulated mass of the dark matter halo depends on which mass
model we are using.

The NFW DM profile emerged as the universal density profile for
the dark matter halo from the simulations of $\Lambda$CDM model.
The density profile depends on the critical density required to
flatten the universe and therefore it is determined uniquely by
the Hubble parameter $H$.  The profile correctly gives a constant
rotation velocity at large radii for most galaxies.  However,
recent observation of the rotation curves from the late-type LSB
galaxies reveals that the center of these galaxies contains an
approximately constant density core better parametrized by the ISO
DM profile~\cite{de Blok:2002tg}.  Among the LSB galaxies in
Ref.~\cite{de Blok:2002tg}, U4173, U4325, U3371, DDO185, DDO47,
DDO64, U1281, DDO52, IC2233 are the galaxies which the NFW profile
cannot be used to fit with the observational data of the rotation
curves around the core region without stretching it to unrealistic
parameters, namely too large asymptotic rotation velocity at large
radii.  On the other hand, the ISO profile can fit well to most
LSB galaxies implying that these galaxies might actually have a
constant core region.

The contrast between simulation results from the $\Lambda$CDM model
and observational data is known as the core-cusp problem~(see
Ref.~\cite{cc} for an excellent review).  While the $\Lambda$CDM
model supports a cuspy rotation curve, observational data of the
late-type LSB galaxies prefer the  mass model with a constant core.
To explore the effects of the chameleonic fifth force on each kind of DM
halo, we consider two popular mass models, NFW and ISO, and finally
in a parametrized~(PM) model.

\subsection{NFW halo}

For the NFW profile, the DM density is given by
\begin{eqnarray}
\rho_{\text{NFW}}(r) =
\frac{\rho_0}{\frac{r}{a}\left(1+\frac{r}{a}\right)^2},
\end{eqnarray}
where $\rho_0$ and $a$ are the characteristic density and the scale
radius of the halo respectively.  Although we will use the NFW
profile and effective acceleration from the chameleonic fifth force
to calculate the circular velocity, we must know the characteristic
density of the NFW profile. In order to obtain the $\rho_0$ of NFW
profile, we consider the accumulated mass
\begin{equation}
M_{NFW}(r) =
V_{200}^2\left(\frac{\text{ln}(1+(cr/r_{200}))-(cr/r_{200})/(1+(cr/r_{200}))}{\text{ln}(1+c)-c/(1+c)}\right)\frac{r_{200}}{G},
\end{equation}
leading to the circular velocity in the absence of the chameleon
as
\begin{equation}
v_{c}(r) =
V_{200}\sqrt{\frac{1}{x}\frac{\text{ln}(1+cx)-cx/(1+cx)}{\text{ln}(1+c)-c/(1+c)}},
\end{equation}
where $V_{200}$ is the circular velocity at virial radius
$(r_{200})$ , $x = r/r_{200}$ is the radial distance in unit of
the virial radius and $c = r_{200}/a$ is the concentration
parameter.

The characteristic density of the NFW profile is then given by
\begin{equation}
\rho_0 = \frac{V_{200}^2}{4\pi G a^2}\left(\frac{c}{\text{ln}(1+c)-c/(1+c)}\right).
\end{equation}
And from the virial mass $(m_{200})$, we obtain the virial radius
and the characteristic radius respectively:
\begin{eqnarray}
r_{200} &=& \frac{V_{200}}{10 H}, \\
a &=& \frac{r_{200}}{c}
\end{eqnarray}
where $H$ is the Hubble parameter.

\begin{figure}[htp]
\centering
\includegraphics[width=0.40\textwidth]{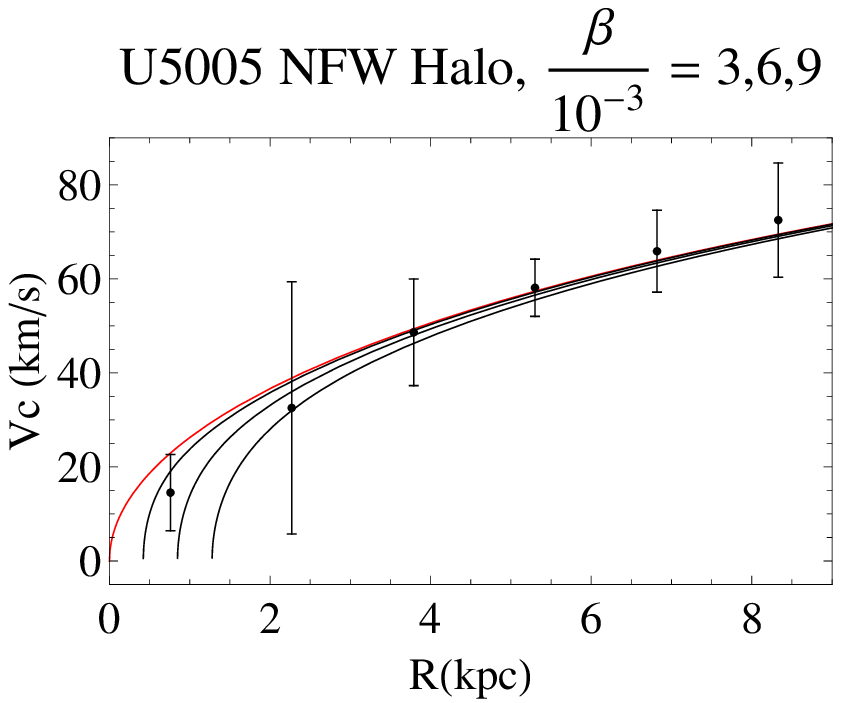} \hfill
\includegraphics[width=0.40\textwidth]{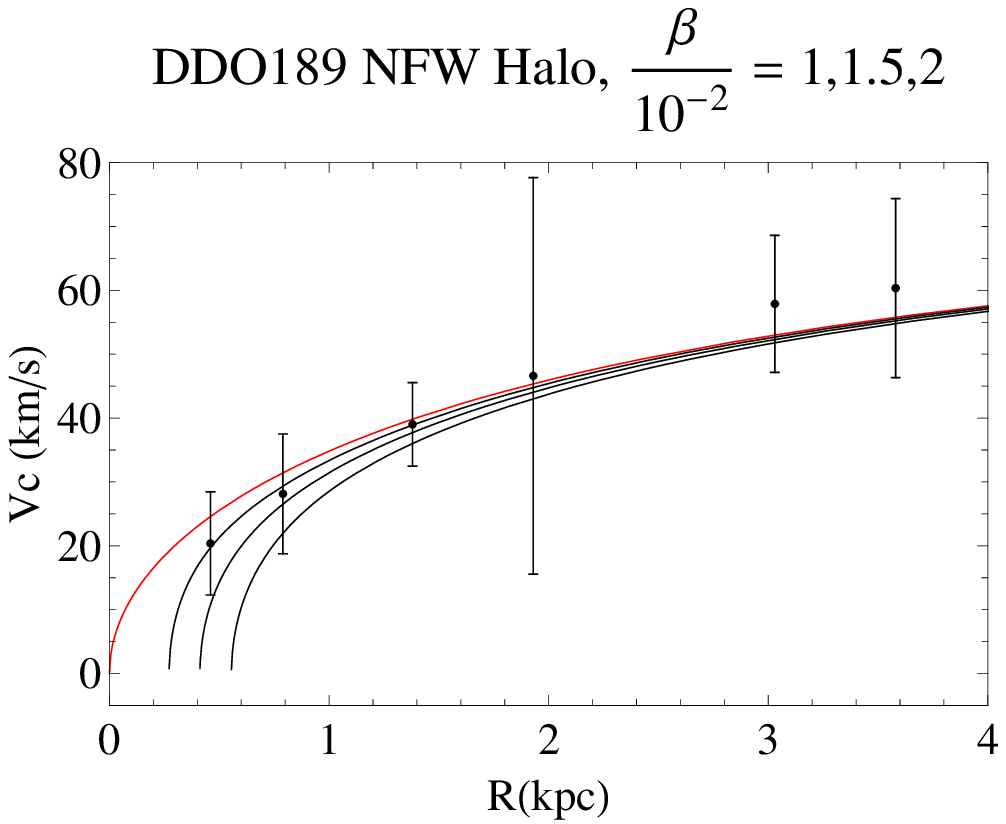}\\
\caption[U4325v]{Rotation curves of U5005, DDO189 galaxy around the
core region for varying $\beta$. The red lines represent the
rotation curve of the galaxy with NFW profile without the chameleon.
}\label{fig1}
\end{figure}

For the late-type low surface brightness~(LSB) galaxies,
dominating gravitational element is the dark matter halo.  It is
thus the most efficient to study effects of chameleonic fifth force on
the rotation curves of the LSB galaxies.
 Using values of $V_{200}$, $c$ from Ref.~\cite{de Blok:2002tg} and $H=72$ km/s/Mpc, we
can produce the rotation curves for certain LSB galaxies shown in
Fig.~\ref{fig1}.

\begin{figure}[htp]
\centering
\includegraphics[width=0.40\textwidth]{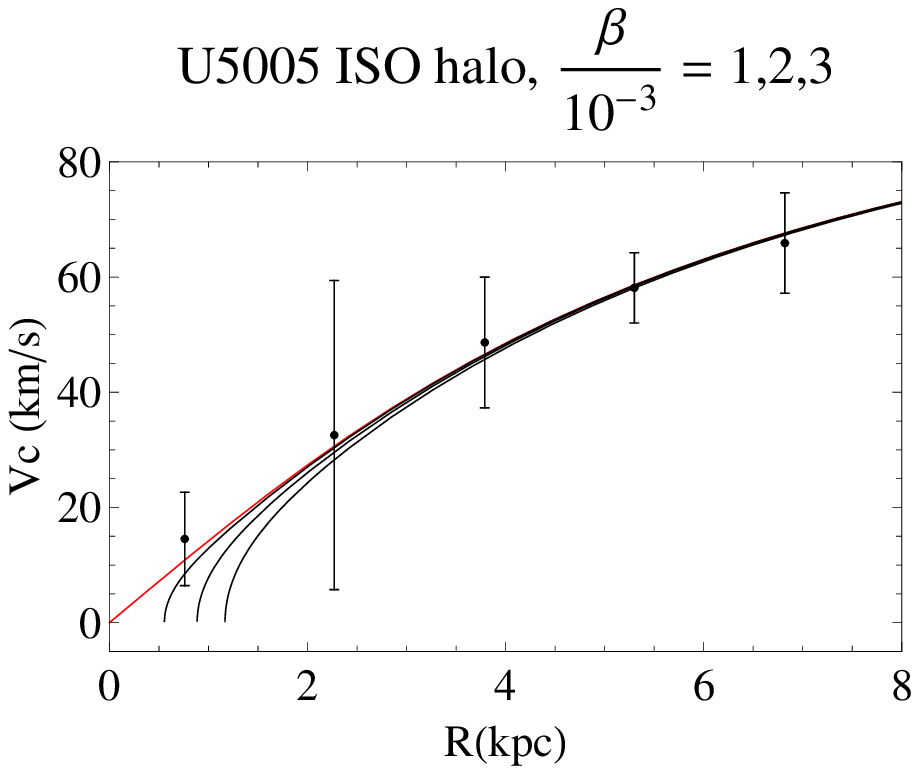} \hfill
\includegraphics[width=0.40\textwidth]{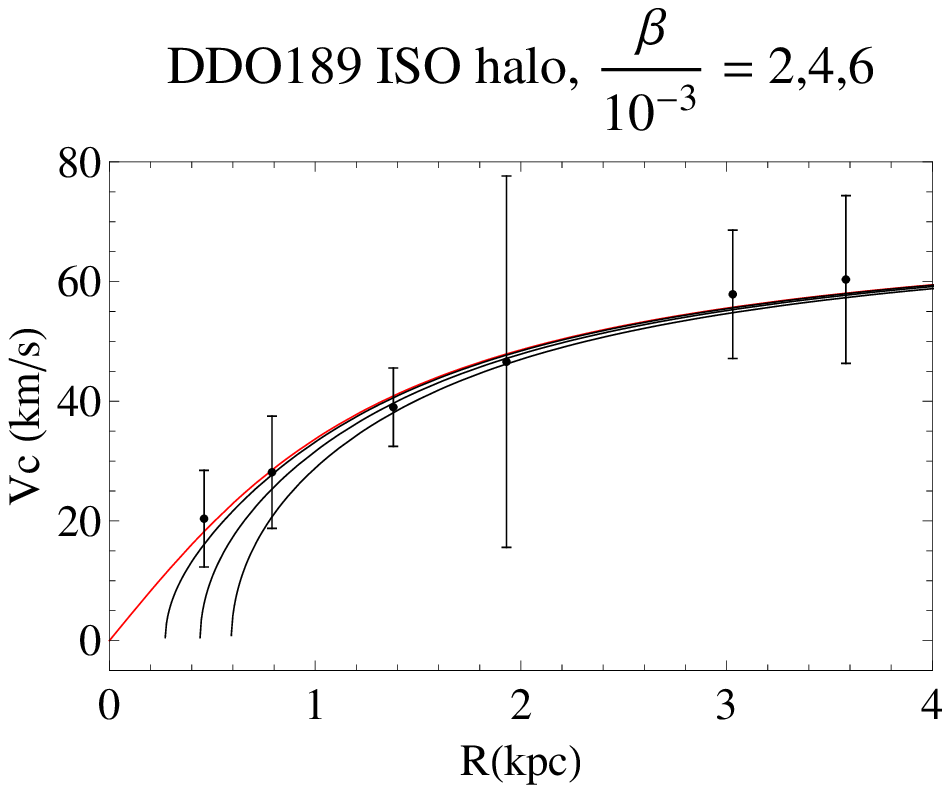}\\
\includegraphics[width=0.40\textwidth]{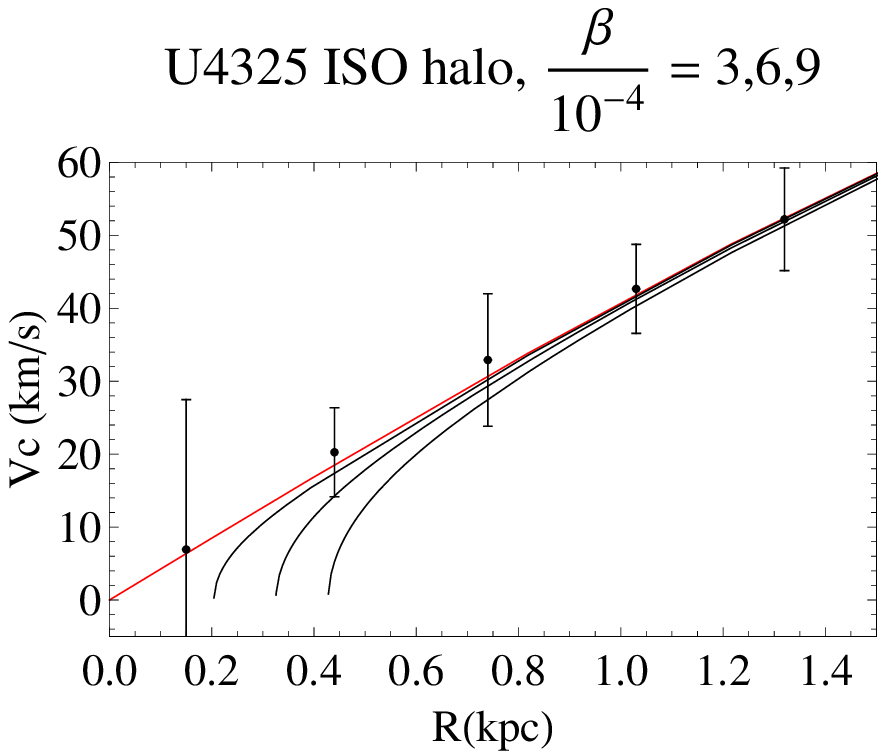} \hfill
\includegraphics[width=0.40\textwidth]{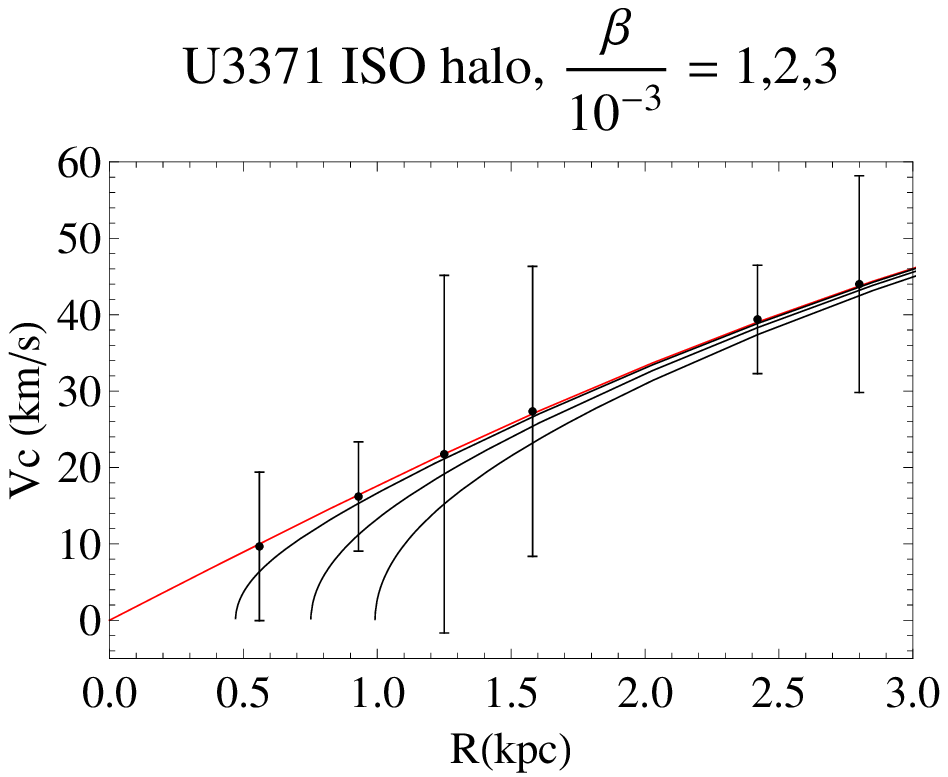}\\
\caption[ISO]{Rotation curves of U5005,~DDO189,~U4325,~U3371 galaxy
in the core region for varying $\beta$. The red lines represent
rotation curves of the galaxy with ISO profile without the
chameleon. }\label{fig2}
\end{figure}

\subsection{ISO Halo}

For a spherical pseudo-isothermal halo, the DM density profile is
assumed to be
\begin{eqnarray}
\rho_{\text{ISO}}(r)& = &
\frac{\rho_c}{1+\left(\frac{r}{R_{c}}\right)^2},
\end{eqnarray}
where $\rho_c$ and $R_{c}$ are the central density and core radius
of the halo respectively.  The rotation curve from this density
profile in the absence of the chameleon is
\begin{eqnarray}
v_{c}(r) & = & \sqrt{4\pi G \rho_{c}R_{c}^{2}
\left(1-\frac{R_{c}}{r}\arctan(\frac{r}{R_{c}})\right)}.
\end{eqnarray}
The best-fit parameters $\rho_{c},R_{c}$ and processed rotation
curves of some LSB galaxies are given in Ref.~\cite{de
Blok:2002tg}. We use these parameters to generate rotation curves
with varying chameleon-matter coupling $\beta$ as are shown in
Fig.~\ref{fig2}.

\subsection{The parametrized model}

\begin{figure}[htp]
\centering
\includegraphics[width=0.40\textwidth]{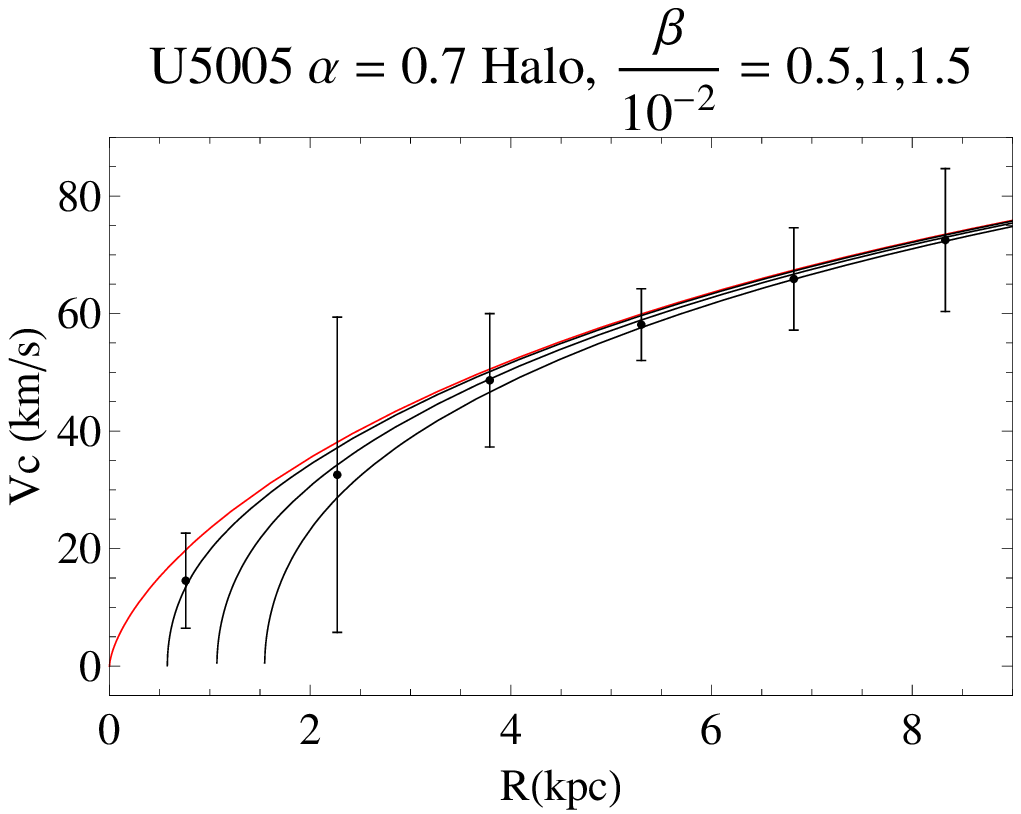} \hfill
\includegraphics[width=0.40\textwidth]{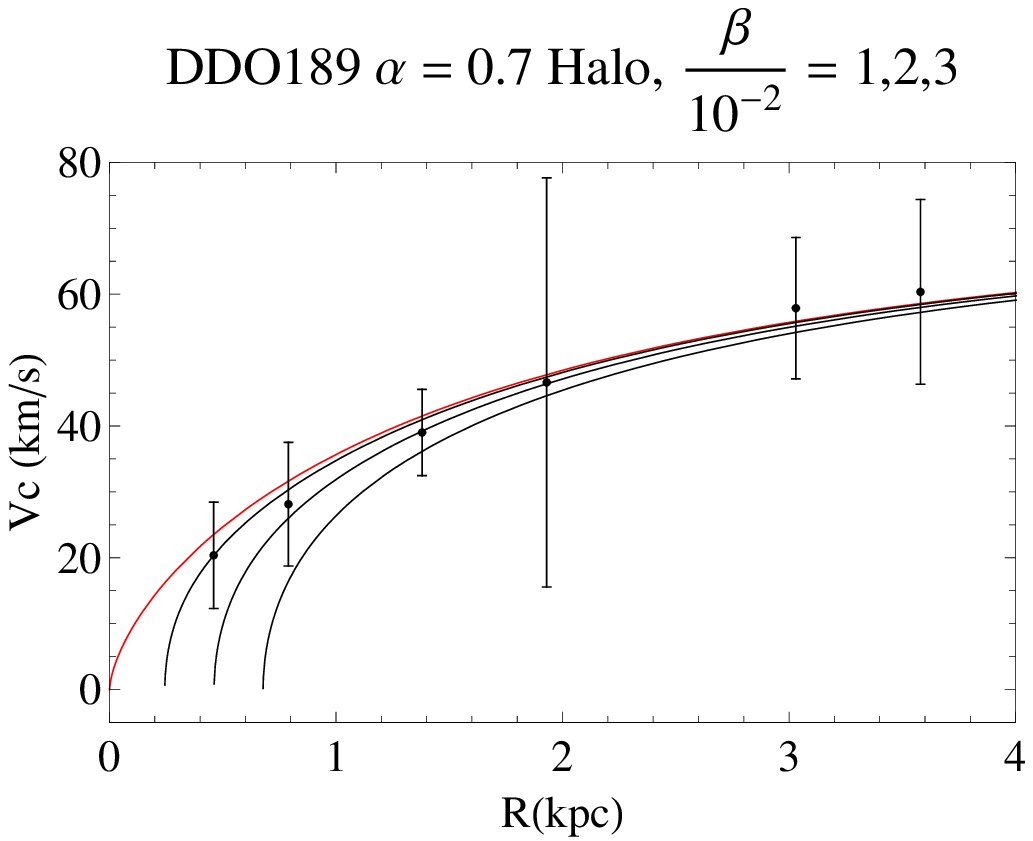}\\
\includegraphics[width=0.40\textwidth]{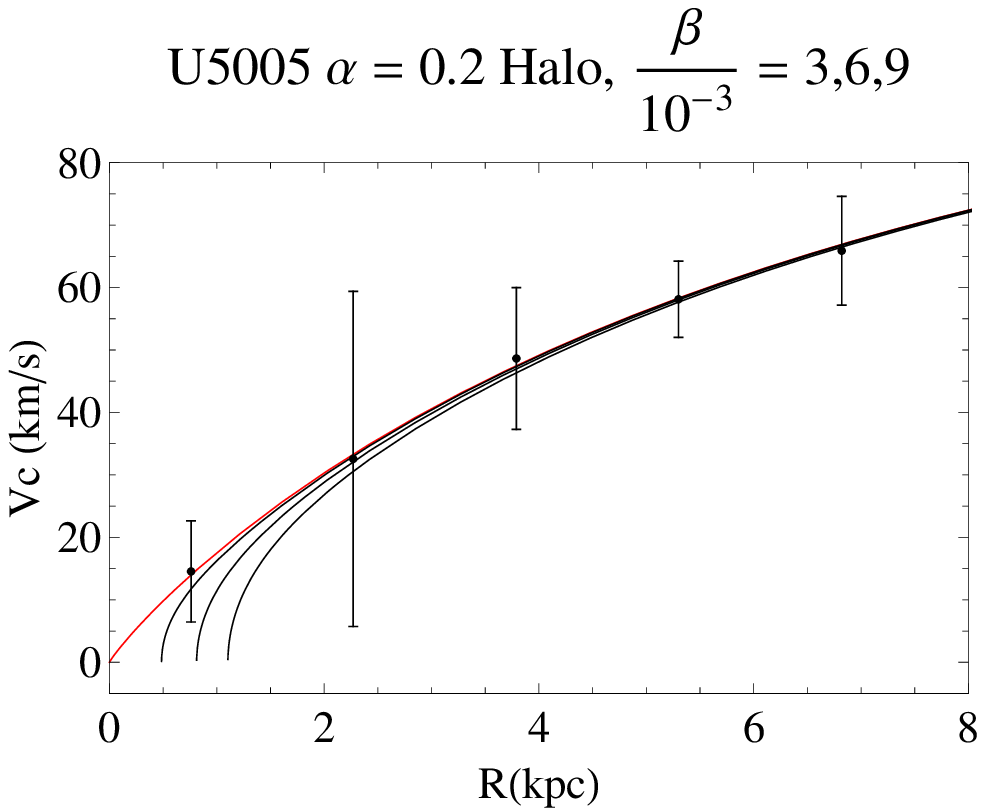} \hfill
\includegraphics[width=0.40\textwidth]{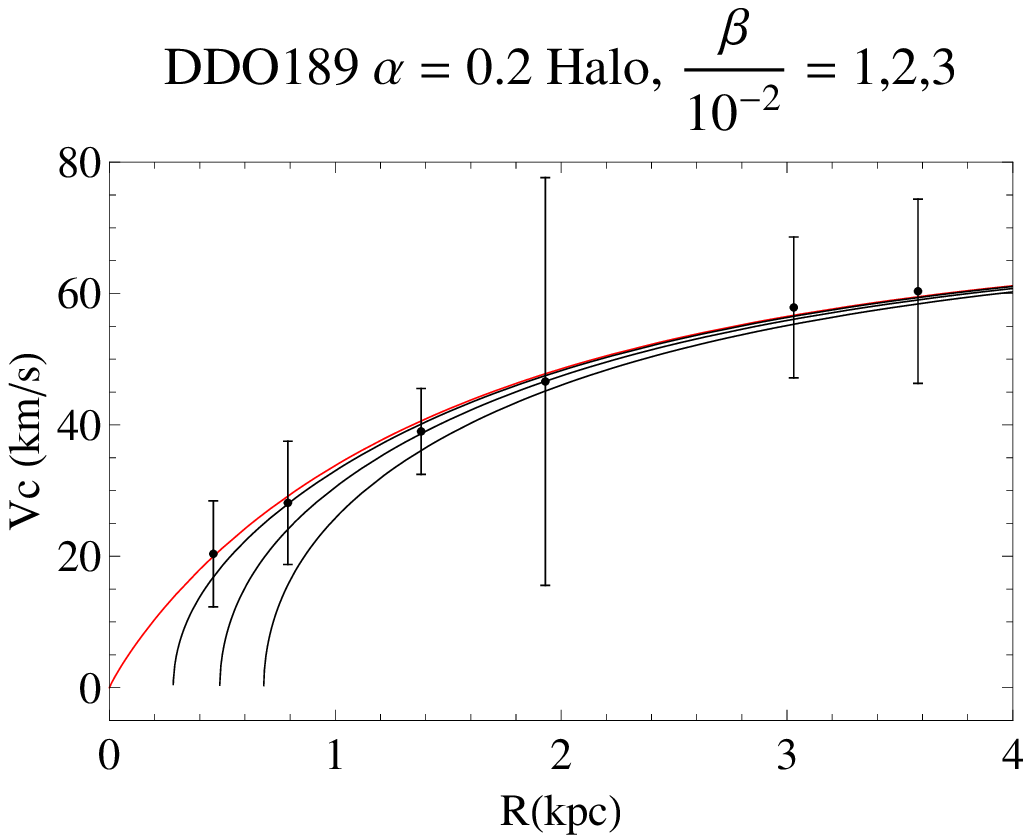}\\
\includegraphics[width=0.40\textwidth]{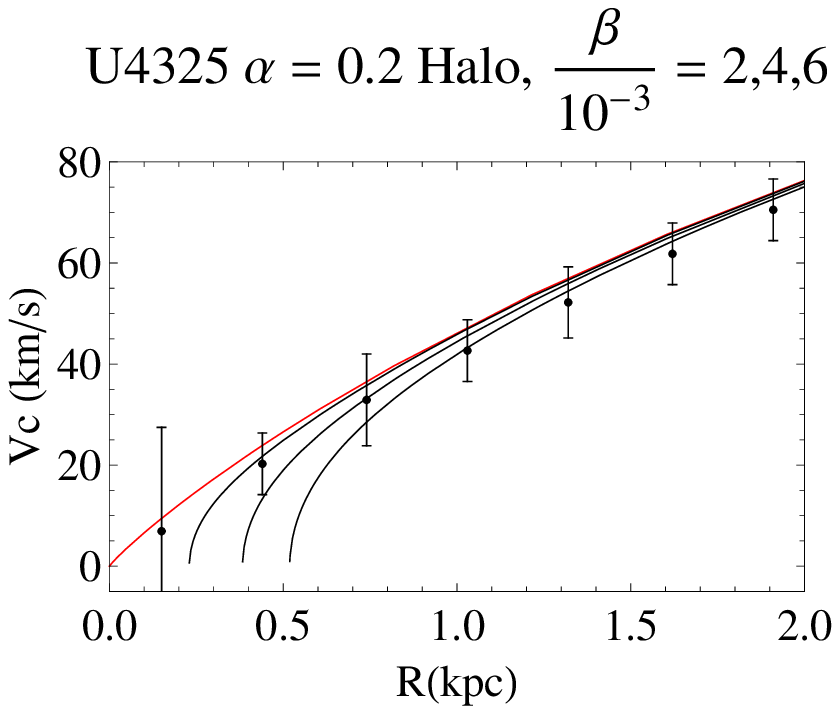} \hfill
\includegraphics[width=0.40\textwidth]{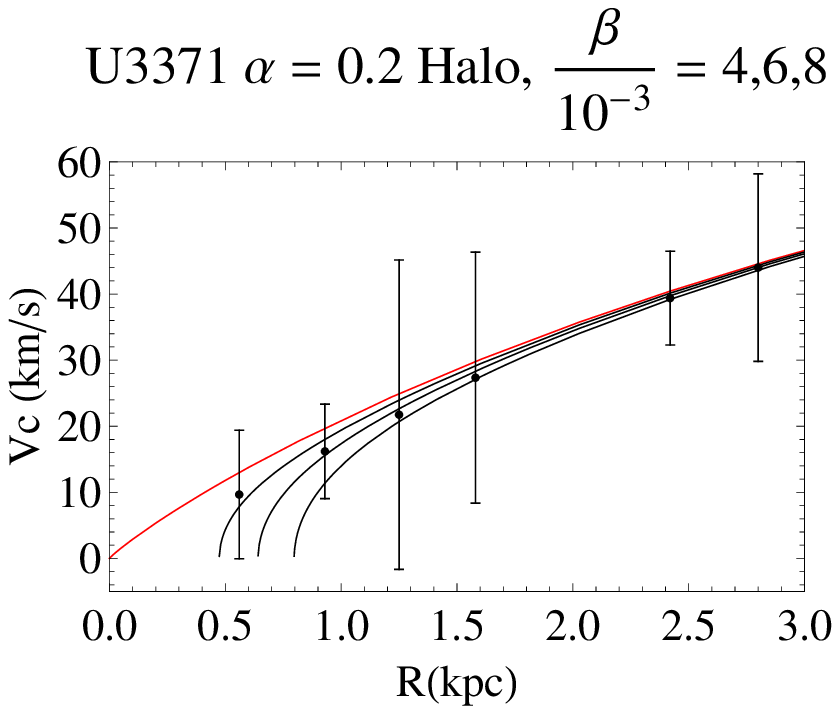}\\
\caption[PM]{Rotation curves of U5005, DDO189, U4325, U3371 galaxy
around the core region for varying $\beta$.  U4325 and U3371 cannot
be fit to the PM model with $\alpha = 0.7$ without making the
rotation velocity unrealistically large.  The red lines represent
the rotation curves of the galaxy with the PM profile without the
chameleon. }\label{fig3}
\end{figure}

We can generalize the density profile of the NFW model to be
\begin{eqnarray}
\rho_{\text{PM}}(r) =
\frac{\rho_0}{(\frac{r}{r_{s}})^{\alpha}\left(1+\frac{r}{r_{s}}\right)^{3-\alpha}},
\end{eqnarray}
where $\alpha$ is a parameter which takes value of 1 for the NFW
profile.  This profile correctly reproduces $\rho \propto r^{-3}$
in the large radius limit.  The circular rotation velocity in the
absence of chameleon is then
\begin{eqnarray}
v_c(r) & = & \sqrt{\frac{4\pi
G\rho_{0}}{3-\alpha}r^{2}\left(\frac{r}{r_{s}}\right)^{-\alpha}{}_{2}F_{1}(3-\alpha,3-\alpha,4-\alpha,-r/r_{s})}
\end{eqnarray}
where ${}_{2}F_{1}$ is the hypergeometric function.  Rotation
curves of U5005, DDO189 with $\alpha = 0.2, 0.7$ and U4325, U3371 with $\alpha = 0.2$ for varying
$\beta$ are shown in Fig.~\ref{fig3}.

\subsection{Analytic approximation of the rotation curve}

An excellent approximation of the rotation curve can be obtained
analytically according to Eqn.~(\ref{genfor}).  Demanding that
$\phi'(r_{max})=0~(i.e. ~\gamma = 0)$ at the galactic boundary, we
obtain $(\phi^{\prime}r^{2}\vert_{r=0}) = -\frac{\beta}{4 \pi
M_{pl}}M(r_{max})$.  The analytic approximation of the acceleration
from the fifth force is then given by
\begin{equation}
a = \frac{\beta}{M_{Pl}}\phi'(r) \simeq  -\frac{1}{4 \pi r^{2}
B(r)}\left(\frac{\beta}{M_{Pl}}\right)^{2}(M_{0} - M(r)).
\label{papp}
\end{equation}
The galactic mass $M_{0}$ is again defined to be $M(r_{max})$. The
force pushes outwardly when the chameleon gradient is negative.
 It effectively reduces the rotation velocity of matter in the galaxy.
 For a generic DM halo which is far from undergoing a gravitational
collapse, $B(r)\simeq 1$. The acceleration from the fifth force is
thus determined mostly by the mass profile and it is proportional to
$\beta^{2}$. Substituting into Eqn.~(\ref{rotc}), the resulting
rotation curves agree extremely well with the numerical results.


\section{Results and Discussions}

For NFW profile, we choose to explore the effects of chameleon in
U5005 and DDO189 galaxy since the data can be fit without assuming
too large circular rotation velocities.  As we can see from
Fig.~\ref{fig1}, the chameleonic fifth force is larger in the core
region. The rotation velocity is reduced as the fifth force drives
the matter in the outward direction.  For U5005, turning the value
of $\beta > 6\times 10^{-3}$ gives the rotation curve cuspier, so
much that it misses the error bar of the innermost data point. For
DDO189, $\beta > 1.75\times 10^{-2}$ results in the bad fit of the
innermost data point.  For statistical analysis, we use U5005 with
the number of degree of freedom $N = 11-3 = 8$ and establish a
constraint on the upper bound of matter coupling $\beta < 1\times
10^{-3}$ at 95 \% C.L. DDO189 gives less stringent constraint.

\begin{figure}[htp]
\centering
\includegraphics[width=0.45\textwidth]{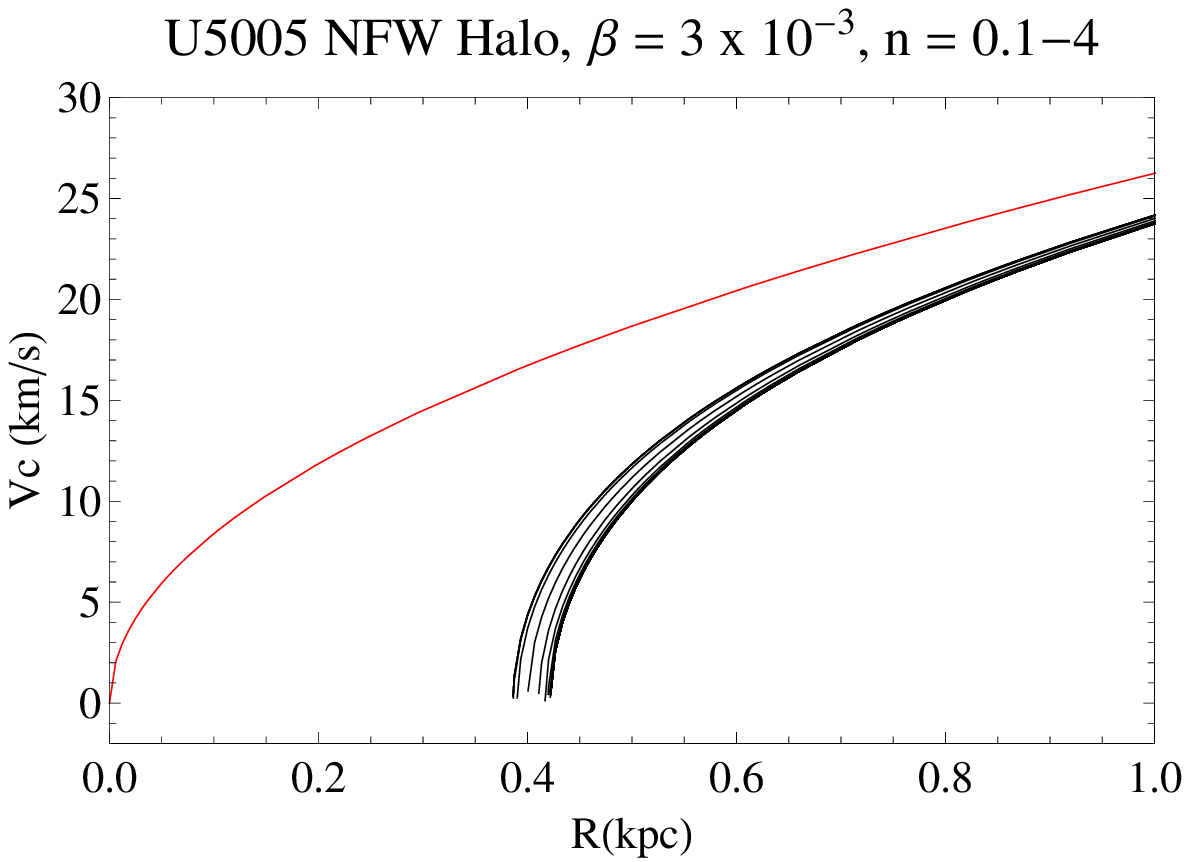} \hfill
\includegraphics[width=0.45\textwidth]{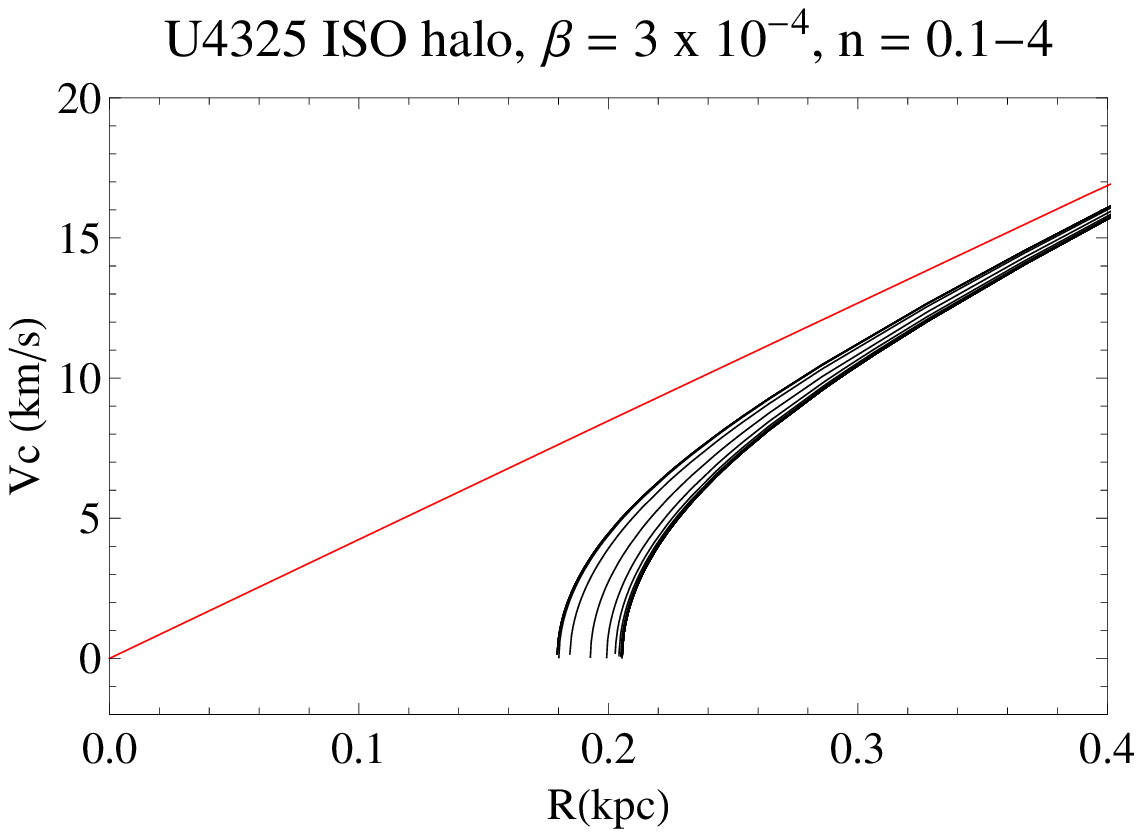}\\
\caption[varyn]{Rotation curves of galaxy U5005, U4325 for the
chameleonic self potential $V(\phi) = M^{n+1}/\phi^{n}$ with $n =
0.1 - 4$~(left to right) using NFW, ISO profile.  The red lines
are the rotation curves without the chameleon. }\label{fign}
\end{figure}

For ISO profile, we present rotation curves for U5005, U4325, U3371,
DDO189 galaxies with varying $\beta$.  The strongest constraint on
the coupling $\beta$ is from U4325 where $\beta > 3\times 10^{-4}$
results in a large deviation of the rotation curve from the
innermost data point. This is shown in Fig.~\ref{fig2}.  For
statistical analysis, we use U4325 with the number of degree of
freedom $N = 16-3 = 13$ and establish a constraint on the upper
bound of matter coupling $\beta < 1\times 10^{-3}$ at 95 \% C.L.
Other galaxies give less stringent constraints.

\begin{table}[hbt]
\centering
\begin{tabular}{| l | c |}
\hline
LSB galaxy & upper bound on $\beta$ at 95 \% C.L. \\[0.5ex]
\hline
U5005 (NFW) & $6\times10^{-3}$  \\[0.5ex]
U5005 (ISO) & $2\times 10^{-3}$  \\[0.5ex]
U5005 (PM $\alpha = 0.2$) & $6\times 10^{-3}$  \\[0.5ex]
U5005 (PM $\alpha = 0.7$) & $9\times 10^{-3}$  \\[0.5ex]\hline
DDO189 (NFW) & $1.75\times 10^{-2}$  \\[0.5ex]
DDO189 (ISO) & $4.8\times 10^{-3}$  \\[0.5ex]
DDO189 (PM $\alpha = 0.2$) & $1.75\times 10^{-2}$  \\[0.5ex]
DDO189 (PM $\alpha = 0.7$) & $1.85\times 10^{-2}$  \\[0.5ex]\hline
U4325 (ISO) & $1\times 10^{-3}$  \\[0.5ex]
U4325 (PM $\alpha = 0.2$) & $5.4\times10^{-3}$  \\[0.5ex]\hline
U3371 (ISO) & $2.7\times 10^{-3}$  \\[0.5ex]
U3371 (PM $\alpha = 0.2$) & $9.5\times10^{-3}$  \\[0.8ex]
\hline
\end{tabular}
\caption{Constraints on the matter-chameleon coupling constant from
the LSB galaxies.}\label{tab}
\end{table}

The PM mass model can fit with the rotation curve data better than
the NFW in the core region for $\alpha < 1$ since the cusp is less
steep. For $\alpha > 1$, the quality of fit is worse, therefore we
consider the PM mass model only in $\alpha < 1$ cases.  This is
shown in Fig.~\ref{fig3}.  Again, the chameleonic fifth force makes
the cusp steeper, resulting in the worse fitting.  For U5005,
DDO189, large deviations of the rotation curve from the innermost
data point occur when $\beta > 1,2\times 10^{-2}$ for $\alpha = 0.7$
respectively. For statistical analysis, we use U4325 with the number
of degree of freedom $N = 16-3 = 13$ and establish a constraint on
the upper bound of matter coupling $\beta < 5.4\times 10^{-3}$ at 95
\% C.L. for $\alpha = 0.2$. For $\alpha = 0.7$, U5005 is used with
$N = 11-3 = 8$, the upper bound on the matter coupling becomes
$\beta < 9\times 10^{-3}$. The strongest constraints for each DM
halo are summarized in Table~\ref{tab}. Numerical studies show that
the rotation curves change very slightly~(cuspier for larger $n$)
for the self potential with $n=0.1-4$ as shown in Fig.~\ref{fign}.
The constraints are thus mostly the same for the power-law potential
with $n=0.1-4$~(only fractionally stronger).

Observe that the rotation curves are made {\it cuspier} for every
galaxy in the presence of the chameleon, regardless of the mass
model of the DM halo. The parameters of each mass model need to be
fit to the rotation curve in the large radii region and they will be
the same as the best fit in the absence of the chameleon. Deviation
of the rotation curves in the presence of the chameleon will appear
in the core region where the effect of the chameleonic fifth force
is the most distinctive.

\section{Conclusions}

Generically, a chameleon scalar field is designed to interact with
matter through a conformal coupling so that its mass varies with
the matter density locally.  The chameleon becomes massive when
the local density is high and the modification of gravity from the
modified Yukawa potential is thus negligible.  The chameleon
mechanism makes the scalar field evade the constraints on the
violation of Equivalence Principle and the fifth force from the
Earth bound experiments.

In the situation where matter density varies, the chameleon obtains
a profile with spatial variation, generating a pressure profile
within the matter.  The pressure gradient force is actually the
fifth force induced by the chameleon-matter coupling.  We
demonstrate that for a sufficiently large massive object such as a
galaxy, the non-singular boundary condition leads to stringent but
yet unphysical constraint on the matter-chameleon coupling~(roughly
of order $\beta < 10^{-7}$).  We argue that the physical chameleon
profile in general inevitably develop a singularity at the center.
The chameleon profile with central singularity leads to a
significantly large chameleonic fifth force close to the center
region. In a presence of the galactic DM halo, the fifth force
becomes appreciably large so that in the core region of the galaxy,
the force significantly reduces the circular rotation velocity of
the galaxy, resulting in a cuspier rotation curve.

We investigate the effects of the chameleonic fifth force on the
rotation curves of certain late-type LSB galaxies using the NFW,
ISO, and parametrized mass profiles. Constraints on the upper bound
at 95 \% C.L. on the chameleon-matter coupling are established in
Table~\ref{tab}.  The constraints could be as stringent as $\beta <
1\times 10^{-3}$ for the ISO mass model. For a parametrized mass
model with $\alpha = 0.7, 0.2$, the constraints are not as strong,
$\beta < 5-9\times 10^{-3}$.

Analytic approximation of the rotation curve in the presence of the
chameleon is derived with great accuracy.  The chameleonic fifth
force is determined mostly by the mass profile of the DM halo and it
is proportional to the matter coupling $\beta^{2}$. The force is
directed outward since the chameleon gradient is negative. The force
is strongest in the core region and the rotation curve is altered
the most in the region.  The chameleon therefore generates cuspier
rotation curves for every mass model of the DM halo.

\section*{Acknowledgments}

We would like to thank Maneenate Wechakama and Wiphu Rujopakarn for
valuable discussions and suggestion on the LSB galaxy data. P.B. and
S.P. are supported in part by the 90th Year Chulalongkorn
Scholarship and Toray Science Foundation, Japan~(TSF).  P.B. is
supported in part by Thailand Center of Excellence in
Physics~(ThEP).

\appendix

\section{Derivation of the chameleonic fifth force} \label{appa}

From the chameleon-matter action
\begin{equation}
S = \int
d^4x\sqrt{-g}\left(\frac{M_{Pl}^2}{2}R-\frac{\left(\partial\phi\right)^2}{2}-V\left(\phi\right)\right)-\int
d^4x\mathcal{L}_{m}\left(\psi_{m},A^2\left(\phi\right)g_{\mu\nu}\right),
\nonumber
\end{equation}
after varying with respect to the field, $\phi$, we obtain an
equation of motion of the chameleon as
\begin{eqnarray}
\nabla^2\phi = V,_{\phi}-\,\alpha_{\phi}T^{\mu (m)}_{\mu}.
\label{eomApp}
\end{eqnarray}
The above equation must be consistent with the equation from
$\nabla^{\mu} T_{\mu\nu} = 0$. Since the chameleon scalar field
couple with matter, then
\begin{equation}
\nabla^{\mu} T^{(total)}_{\mu\nu} = \nabla^{\mu}
T^{(m)}_{\mu\nu}+\nabla^{\mu} T^{(\phi)}_{\mu\nu} = 0. \label{covT}
\end{equation}
The energy-momentum tensor of the scalar field is
\begin{equation}
T^{(\phi)}_{\mu\nu} = \partial_{\mu}\phi\partial_{\nu}\phi
-g_{\mu\nu}\left(\frac{1}{2}g^{\alpha\beta}\partial_{\alpha}\phi\partial_{\beta}\phi+V(\phi)\right).
\nonumber
\end{equation}
Thus
\begin{eqnarray}
\nabla^{\mu} T^{(\phi)}_{\mu\nu} &=& (\nabla^{\mu}\partial_{\mu}\phi)\partial_{\nu}\phi + \partial_{\mu}\phi(\nabla^{\mu}\partial_{\nu}\phi)-g_{\mu\nu}\nabla^{\mu}\left(\frac{1}{2}g^{\alpha\beta}\partial_{\alpha}\phi\partial_{\beta}\phi+V(\phi)\right),  \nonumber  \\
&=& (\nabla^{\mu}\nabla_{\mu}\phi)\partial_{\nu}\phi + \partial_{\mu}\phi(\nabla^{\mu}\partial_{\nu}\phi)-g_{\mu\nu}(\nabla^{\mu}(\frac{1}{2}\partial_{\alpha}\phi\partial^{\alpha}\phi)+\nabla^{\mu}V(\phi)),  \nonumber  \\
&=& (\nabla^2\phi)\partial_{\nu}\phi + \partial_{\mu}\phi(\nabla^{\mu}\partial_{\nu}\phi) - \partial^{\alpha}\phi\nabla_{\nu}\partial_{\alpha}\phi - V,_{\phi}\partial_{\nu}\phi,  \nonumber  \\
&=& (\nabla^2\phi - V,_{\phi})\partial_{\nu}\phi +
\partial_{\mu}\phi(\nabla^{\mu}\partial_{\nu}\phi) -
\partial^{\alpha}\phi\nabla_{\nu}\partial_{\alpha}\phi,   \nonumber
\end{eqnarray}
where
\begin{eqnarray}
\nabla_{\nu}\partial_{\alpha}\phi &=& \partial_{\nu}\partial_{\alpha}\phi - \Gamma^{\lambda}_{\nu\alpha}\partial_{\lambda}\phi, \nonumber  \\
\nabla^{\mu}\partial_{\nu}\phi &=& g^{\mu\alpha}\partial_{\alpha}\partial_{\nu}\phi - g^{\mu\alpha}\Gamma^{\lambda}_{\alpha\nu}\partial_{\lambda}\phi,   \nonumber   \\
&=& \partial^{\mu}\partial_{\nu}\phi -
g^{\mu\alpha}\Gamma^{\lambda}_{\alpha\nu}\partial_{\lambda}\phi.
\nonumber
\end{eqnarray}
Then
\begin{eqnarray}
\partial_{\mu}\phi(\nabla^{\mu}\partial_{\nu}\phi) - \partial^{\alpha}\phi\nabla_{\nu}\partial_{\alpha}\phi &=& \partial_{\mu}\phi(\partial^{\mu}\partial_{\nu}\phi - g^{\mu\alpha}\Gamma^{\lambda}_{\alpha\nu}\partial_{\lambda}\phi) - \partial^{\alpha}\phi(\partial_{\nu}\partial_{\alpha}\phi - \Gamma^{\lambda}_{\nu\alpha}\partial_{\lambda}\phi),  \nonumber \\
&=& -\partial^{\alpha}\phi\Gamma^{\lambda}_{\alpha\nu}\partial_{\lambda}\phi + \partial^{\alpha}\phi\Gamma^{\lambda}_{\nu\alpha}\partial_{\lambda}\phi, \nonumber  \\
&=& 0.   \nonumber
\end{eqnarray}
Therefore
\begin{eqnarray}
\nabla^{\mu} T^{(\phi)}_{\mu\nu} = (\nabla^2\phi -
V,_{\phi})\partial_{\nu}\phi.     \nonumber
\end{eqnarray}
From the equation of motion of the chameleon (Eqn.~(\ref{eomApp})),
we obtain
\begin{eqnarray}
\nabla^{\mu} T^{(\phi)}_{\mu\nu} &=& -\alpha_{\phi}T\partial_{\nu}\phi,  \nonumber  \\
 &=& \alpha_{\phi}\rho\partial_{\nu}\phi.      \nonumber
\end{eqnarray}
From Eqn.~(\ref{covT}), the covariant derivative of the matter is
then
\begin{eqnarray}
\nabla^{\mu} T_{\mu\nu}^{(m)} &=&
-\alpha_{\phi}\rho\partial_{\nu}\phi.   \nonumber
\end{eqnarray}
For non-relativistic fluid and approximately flat spacetime, the LHS
becomes
\begin{eqnarray}
\rho(\partial_{t} v^i + \vec v\cdot \vec\nabla v^i) +\partial_i P
&=&  -\alpha_{\phi}\rho\partial_{i}\phi.    \nonumber
\end{eqnarray}
We approximate the matter to be pressureless while the gradient has
only the radial direction due to spherical symmetry ($\vec v \cdot
\vec \nabla = 0$). Therefore
\begin{eqnarray}
\rho(\partial_{t} v^i) &=& -\alpha_{\phi}\rho\partial_{i}\phi,
\nonumber
\end{eqnarray}
then
\begin{eqnarray}
\vec a = -\alpha_{\phi}\vec \nabla\phi.    \nonumber
\end{eqnarray}
This is the acceleration due to the fifth force from the chameleon
acting on the matter.
Therefore, if we consider only the effects on the matter in presence
of gravity , we will obtain the circular velocity of matter within
the galaxy as
\begin{eqnarray}
v_c(r) = \sqrt{\frac{G M(r)}{r} + \alpha_{\phi}(\partial_r\phi)r}.
\nonumber
\end{eqnarray}

\end{document}